%% file: rjr.tex
\documentclass[a4paper,fleqn,10pt]{scrartcl}
\pdfoutput=1
\input{preample}

\hypersetup{
  pdfauthor={Nick Baberuxki, Christian T Preuss, Daniel Reichelt,
             Steffen Schumann}, 
  pdftitle={Resummed predictions for jet-resolution scales in multijet production in $e^+e^-$ annihilation}
}

\preprint{MCNET-19-29 \\FERMILAB-PUB-19-639-T}

\author[1]{Nick Baberuxki}
\author[2]{Christian T Preuss}
\author[1,3]{Daniel Reichelt}
\author[1]{Steffen Schumann}

\affil[1]{Institut f{\"u}r Theoretische Physik,
  Georg-August-Universit{\"a}t G{\"o}ttingen, D-37077 G{\"o}ttingen, Germany}

\affil[2]{School of Physics and Astronomy, Monash University, Clayton VIC-3800, Australia}

\affil[3]{Fermi National Accelerator Laboratory, P. O. Box 500, Batavia, IL 60510, USA}

\title{Resummed predictions for jet-resolution scales in multijet production in $e^+e^-$ annihilation}

\newcommand{\Nc}{\ensuremath{N_{\text{C}}}}
\newcommand{\TR}{\ensuremath{T_\text{R}}}
\newcommand{\CA}{\ensuremath{C_\text{A}}}
\newcommand{\CF}{\ensuremath{C_\text{F}}}
\newcommand{\nf}{\ensuremath{n_\text{f}}}
\newcommand{\Rs}{\ensuremath{R_{\text{s}}}}
\newcommand{\lambdas}{\ensuremath{\lambda_{\text{s}}}}
\newcommand{\SU}{\text{SU}}
\newcommand{\gS}{\ensuremath{g_\text{s}}}
\newcommand{\alphaS}{\ensuremath{\alpha_\text{s}}}
\newcommand{\alphaSzero}{\ensuremath{\alpha_{\text{s},0}}}
\newcommand{\imag}{\text{i}}
\DeclareUnicodeCharacter{2212}{-}

\begin{document}
\maketitle

\begin{abstract}
  We present for the first time resummed predictions at $\mathrm{NLO+NLL}^\prime$ accuracy
  for the Durham jet-resolution scales $y_{n,n+1}$ in multijet
  production in $e^+e^-$ collisions. Results are obtained using an implementation
  of the well known \Caesar formalism within the \Sherpa framework. For the 4-, 5- and 6-jet resolutions
  we discuss in particular the impact of subleading colour contributions and compare
  to matrix-element plus parton-shower predictions from \Sherpa and \Vincia.
\end{abstract}		

\tableofcontents
\clearpage

\section{Introduction}

Jet-production processes provide direct means to investigate the dynamics of the strong interaction.
Especially multijet production poses a severe multi-scale problem complicating theoretical analyses. 
Besides integrated jet rates, differential distributions of jet-resolution scales give insight into
the mechanism of jet production for a given jet-clustering algorithm. While for well-separated energetic
jets a fixed-order QCD estimation might be adequate, in particular for small jet-resolutions this
is bound to fail and all-orders resummation techniques need to be employed to provide reliable
theoretical predictions. These account for soft and collinear QCD radiation that dominates the
emission spectrum.

While jet rates are of phenomenological importance both at lepton and hadron colliders, we here
focus on the resummation of jet-resolution scales in multijet production in $e^+e^-$ annihilation.
In particular, we consider the Durham jet algorithm~\cite{Catani:1991hj}, where the jet-resolution
parameter is determined by
\begin{equation}\label{eq:jet_def}
y_{ij} = \frac{2\min(E^2_i,E^2_j)}{Q^2}(1-\cos\theta_{ij})\,.
\end{equation}
Here, $E_{i,j}$ denote the energies of (pseudo) particles $i$ and $j$, $\theta_{ij}$ the angle between
them, and $Q^2$ the squared centre-of-mass energy. This measure is used to
successively cluster particles into jet objects. In what follows, we are interested in the resolution
scales where an $(n+1)$-jet final state is clustered in an $n$-jet final state, i.e., where the emission of
an additional jet off the $n$-jet configuration gets resolved. More precisely,
we consider the differential $3\to 4$, $4\to 5$ and $5\to 6$ jet resolutions at
next-to-leading logarithmic accuracy, i.e., resumming leading logarithms (LL) of type $\alphaS^kL^{k+1}$
and next-to-leading logarithms (NLL) $\alphaS^kL^{k}$ with $L\equiv -\ln(y_{n,n+1})$ appearing in the exponent
of the observable's cummulant distribution. For many observables in $e^+e^-$ annihilation, there
exist results at NLL, NNLL or even N$^3$LL order matched to exact NLO or NNLO QCD matrix elements, see
for instance~\cite{Catani:1992ua,Banfi:2000si,Schwartz:2007ib,Becher:2008cf,Abbate:2010xh,Monni:2011gb,Becher:2012qc,Abbate:2012jh,Larkoski:2014uqa,Banfi:2014sua}.
However, these are limited to two- or three-jet final states. Here, we are in particular exploring high-multiplicity
processes, i.e., jet emission from $4$- and $5$-jet matrix elements that feature Born processes with
non-trivial colour configurations. For the three-jet resolution $y_{23}$, NLL resummed predictions were
calculated in \cite{Banfi:2001bz}; NNLL+NNLO results were presented in~\cite{Banfi:2016zlc}. This
observable in particular has been used in extractions of the strong coupling constant $\alphaS$ from
LEP data, see for instance~\cite{Dissertori:2009qa,Bethke:2008hf,Dissertori:2009ik}. Recently, in
Ref.~\cite{Verbytskyi:2019zhh}, the extraction of $\alphaS$ from a simultaneous fit of the two- and
three-jet rates has been presented. However, in this study the three-jet rate has been considered at
fixed-order NNLO QCD without resummation of logarithmic enhanced terms. Up to now, for Durham
three- and higher jet rates only the next-to-double-leading logarithms of order $\alphaS^kL^{2k-1}$ have
been calculated~\cite{Catani:1991hj}. In \cite{Gerwick:2012fw} this has been extended to the generalised
class of $k_T$-type algorithms defined in~\cite{Cacciari:2011ma}. Besides for the extractions of $\alphaS$,
integrated and differential jet rates prove to be very useful for theoretical validation of parton-shower
algorithms~\cite{Gerwick:2012fw,Webber:2010vz}. Corresponding experimental measurement data from the LEP
experiments, see for example~\cite{Heister:2003aj,Abbiendi:2004qz}, form an integral ingredient to
event generator tuning efforts, as they provide handles to constrain the parameters of phenomenological
hadronisation models, see Ref.~\cite{Buckley:2011ms} for a review. 

To achieve NLL resummation for jet resolutions, we rely on an independent implementation of
the \Caesar formalism~\cite{Banfi:2003je,Banfi:2004yd} in the \Sherpa~\cite{Gleisberg:2008ta,Bothmann:2019yzt}
event-generator framework, presented in~\cite{Gerwick:2014gya}. A brief introduction to the \Caesar
approach and details on the implementation of automated NLL resummation in the \Sherpa framework
are discussed in Sec.~\ref{sec:caesar_sherpa}. This includes a brief presentation of the approach
used to match our resummed predictions to exact QCD matrix elements. In Sec.~\ref{sec:resummed_results},
we give specific algorithmic details on the actual resummation of jet-resolution scales and
discuss the validation of the soft function and the multiple-emission contribution in particular. 

In Sec.~\ref{sec:results}, we present our resummed predictions for jet production off $3-,4-$ and $5-$jet
final states in $e^+e^-$ annihilation. To this end we evaluate the jet resolutions $y_{34}$, $y_{45}$ and
$y_{56}$ at NLL accuracy, matched to exact $4-,5-$ and $6-$jet NLO QCD matrix elements, respectively.
We give an account of subleading-colour contributions, by comparing our full-colour results to the strict
and an improved large-$\Nc$ approximation, the latter corresponding to what is typically used in QCD parton
showers. Finally, we compare the $\mathrm{NLO}+\mathrm{NLL}^\prime$ results to
matrix-element-plus-parton-shower simulations from \Sherpa and \Vincia, where we
also address the impact of hadronisation corrections on the observables. We
present our conclusions and an outlook in Sec.~\ref{sec:conclusions}.

\section{Semi-automated resummation within the \Sherpa framework}\label{sec:caesar_sherpa}

In Ref.~\cite{Gerwick:2014gya}, an implementation of the \Caesar\ formalism in the form of a plugin to
the \Sherpa event-generator framework has been presented, which we employ for resummation of
Durham jet-resolutions. A particular focus has there been put on validating the colour decomposition
of hard-scattering matrix elements and the colour-insertion operators for multi-parton processes. The
resummation plugin has since been used for NLL resummation of the thrust event-shape variable in
$e^+e^-$, $ep$ and $pp$ collisions~\cite{Gerwick:2014gya}, with results matched to tree-level
real-emission matrix elements using a point-wise local subtraction technique. Recently, the
implementation has also been used to resum the soft-drop thrust shape in $e^+e^-$ annihilation~\cite{Marzani:2019evv}. 

In this section, we briefly review the \Caesar\ formalism and comment on specific implementation
aspects. Details on the application to resummation of jet-resolution scales will then be
provided in Sec.~\ref{sec:resummed_results}.

\subsection{The \Caesar\ formalism in a nutshell}

The \Caesar formalism has originally been presented for final-state resummation in~\cite{Banfi:2003je};
its extension to hadronic initial-states has been worked out in~\cite{Banfi:2004nk,Banfi:2010xy}. 
For an extensive review we refer to~\cite{Banfi:2004yd}.
The method provides all necessary ingredients to perform NLL resummed calculations of recursively
infrared and collinear safe global observables in a largely automated manner. The method is based on the
observation that for the wide class of global event-shape observables, the
resummed cumulative distribution for the observable value $V\leq v$ can, to NLL
accuracy, be expressed in the simple form 
\begin{equation}\label{eq:CAESAR}
  \begin{split}
    \Sigma_\mathrm{res}(v) &= \sum_\delta \Sigma^\delta(v)\,,\,\,\text{where} \\  
    \Sigma_\mathrm{res}^\delta(v) &= \int d\mathcal{B_\delta}
    \frac{\mathop{d\sigma_\delta}}{\mathop{d\mathcal{B_\delta}}} \exp\left[-\sum_{l\in\delta}
      R_l^\mathcal{B_\delta}(L)\right]\mathcal{S}^\mathcal{B_\delta}(L)\mathcal{F}^\mathcal{B_\delta}(L)\mathcal{H}^{\delta}(\mathcal{B_\delta})~,
    \hspace{1cm} L\equiv\ln 1/v
  \end{split}
\end{equation}
where the phase-space integral extends over the Born configurations $\mathcal{B}$ for
each partonic channel $\delta$. Dependence of the various contributions on those will
implicitly be assumed in the following, although labels will be dropped. The
jet-function $\mathcal{H}$ implements kinematic cuts on the Born events and ensures
that only sufficiently hard configurations yield non-zero contributions. In the 
exponent, the collinear radiators for all hard legs $l$ are summed. The function
$\mathcal{F}$ accounts for the effect of multiple emissions, whereas the soft
function $\mathcal{S}$ captures colour correlations. Its logarithmic dependence is defined via
\begin{equation}\label{eq:Sfunc}
  \mathcal{S}(L) = S\left(T(L)\right)\,, \quad \text{with}\quad T(L)=-\frac{1-\frac{2}{a}\lambda}{\pi\beta_0}\,,
\end{equation}
where the parameter $\lambda$ is given by $\lambda=\alphaS(\mu_R^2)\beta_0 L$. 

In \cite{Banfi:2004yd} the functions $R_l$ have been evaluated for the general class
of observables $V$, where the impact of one additional, arbitrarily soft, emission
with momentum $k$ off leg $l$ can be parameterised as
\begin{equation}
  V(k) = d_l(\mu_Q) g_l(\phi_l) \left(\frac{k_{\mathrm{T},l}}{\mu_Q}\right)^a e^{-b_l\eta_l}\,, \label{eq:V(k)}
\end{equation}
i.e., in terms of the transverse momentum $k_{\mathrm{T},l}$, rapidity $\eta_l$, and
azimuth $\phi_l$ of the emission, measured with respect to leg $l$. Here, $\mu_Q$
denotes the, in principle arbitrary, resummation "starting"-scale, to be
distinguished from the centre-of-mass energy $Q$. Independence of the observable
from the unphysical scale $\mu_Q$ implies $d_l \propto \mu_Q^a$. Explicit
results for the radiators can be found in the appendix of~\cite{Banfi:2004yd}.

\subsection{Aspects of automated NLL resummation}\label{sec:automation}

The \Caesar\ formalism is ideally suited for the automation of resummation of appropriate
observables. With the observable parametrisation in terms of $a$, $b_l$, $d_l$
and $g_l$, the radiator functions are known. 
The colour decomposition of the Born process in a
suitable basis and the corresponding soft-gluon correlators, needed for constructing the
soft-function $\mathcal{S}$, are independent of the actual observable to be resummed and can thus
be pre-computed. The required Born-process partial amplitudes can be obtained from an
automated matrix-element generator such as \Comix~\cite{Gleisberg:2008fv} for, in principle,
arbitrary processes. However, the colour-space dimensionality quickly grows with the number
of external partons~\cite{Gerwick:2014gya}, making calculations for high-multiplicity processes
memory-intense. This motivates the construction of optimal, i.e., minimal-dimensional and
orthogonal, bases, see Ref.~\cite{Keppeler:2012ih,Sjodahl:2015qoa,Sjodahl:2018cca}. An observable-dependent
component that needs special attention is the multiple-emission function $\mathcal{F}$, for that
one often has to resort to numerical evaluations, which can, however, be pre-computed and tabulated.
Here, we briefly introduce $\mathcal{S}$ and $\mathcal{F}$ in general terms, before discussing specific
aspects and their validation for the resummation of jet-resolution scales in Sec.~\ref{sec:resummed_results}.

\subsubsection*{The soft function $S$}

The automated generation of the soft-function $S$ as defined in Eq.~(\ref{eq:Sfunc})
for arbitrary multiplicity Born processes has been discussed in detail in Ref.~\cite{Gerwick:2014gya}.
Here, we only want to review some general aspects and briefly discuss the specific colour
spaces we face for the case of multijet production in $e^+e^-$ annihilation.

The functional form of $S$ is given by
\begin{equation}
  \begin{split}
  S(t) = \frac{\expval{e^{-\frac{t}{2}\Gamma^\dagger} e^{\frac{t}{2}\Gamma}}{\mathcal{B}}}{\braket{\mathcal{B}}{\mathcal{B}}}\,,
  \end{split}
\end{equation}
with $\ket{\mathcal{B}}$ the Born-process matrix element and $\Gamma$ the soft anomalous-dimension matrix,
cf.~Eq.~(\ref{eq:softAnomalousDim}). By decomposing the Born matrix element over a colour basis\footnote{We refer to all spanning sets that allow us to represent arbitrary colour structures of processes as bases, although practical choices are usually overcomplete.},
\begin{equation}
\ket{\mathcal{B}} = \sum\limits_{\alpha} \mathcal A_\alpha \ket{b_\alpha}, 
\end{equation}
and making a particular choice $\{\ket{b_\alpha}\}$ for the colour structures of the Born process, we can define a metric 
\begin{equation}
	c_{\alpha\beta} = \braket{b_\alpha}{b_\beta} \label{eq:colour_metric}
\end{equation}
and its inverse $c^{\alpha\beta}$.
This allows us to write
\begin{equation}
  \braket{\mathcal{B}}{\mathcal{B}} = \Tr\left[c H\right]
\end{equation} 
with the hard matrix $H_{\alpha\beta} = \overline{\mathcal A}_\alpha \mathcal A_\beta$ in terms of partial amplitudes $\mathcal A_\gamma$. 
In this notation, the soft function may be written as 
\begin{equation}\label{eq:S_trace}
  S(t) = \frac{\Tr\left[He^{-\frac{t}{2}\Gamma^\dagger}ce^{-\frac{t}{2}\Gamma}\right]}{\Tr\left[cH\right]}\,,
\end{equation}
with
\begin{equation}
  \Gamma = -2 \sum_{i<j} \boldsymbol{T}_i \boldsymbol{T}_j \ln \frac{Q_{ij}}{Q} +
  \imag \pi\sum_{\substack{ij=\\II,FF}} \boldsymbol{T}_i \boldsymbol{T}_j \,. \label{eq:softAnomalousDim}
\end{equation}
Here, the first sum runs over all colour dipoles $ij$, while the second one runs over those consisting only of final-final
or inital-initial configurations, corresponding to the exchange of Coulomb gluons~\cite{Dokshitzer:2005ig}.
Explicit results for $H$, $c$, and $\Gamma$ are known for up to four hard
(coloured) legs; the form of $S$ and the structure of $\Gamma$,
however, holds more generally \cite{Bonciani:2003nt}. 

We obtain the hard matrix $H$ from the matrix-element generator \Comix~\cite{Gleisberg:2008fv} that is part of \Sherpa.
All calculations of colour correlators relevant here contain at least one quark-antiquark pair and are performed in the trace basis. An all-orders trace basis is obtained by connecting each quark with an antiquark in all possible ways and subsequently attaching all gluons in all possible ways. Colour correlators are then evaluated by explicitly employing $\SU(\Nc)$ identities. We note that this slightly differs from the approach in the \ColorFull~\cite{Sjodahl2015colorfull} package, which uses specialised replacements in the trace basis instead.

\clearpage
\begin{longtable}[th!]{lccccc}\toprule
	~ & \multicolumn{5}{c}{\textbf{Dimension}} \\ \midrule
	\textbf{Process} $e^+e^-\to$ & $q\bar q g$ & $q\bar q q \bar q$ & $q\bar q gg$ & $q\bar q q\bar q g$ & $q\bar q ggg$ \\
	\textbf{Colour Space} & 1 & 2 & 3 & 4 & 10 \\
	\textbf{Trace Basis} & 1 & 2 & 3 & 4 & 11 \\
	\textbf{Reduced Trace Basis} & 1 & 2 & 3 & 4 & 10 \\ \bottomrule
	\caption{Colour-space and corresponding bases dimensionalities for all relevant Born channels.}
	  \label{tab:colour_dimensions}
\end{longtable}

One problem arising in the context of trace bases is that these are generally overcomplete for
$n_{q\bar q} + n_g > \Nc$, such that the metric defined in
Eq.~(\ref{eq:colour_metric}) does not possess a unique inverse.
Although this has generally been solved in Ref.~\cite{Gerwick:2014gya}, within
this study we can circumvent it in an even simpler way. The only critical process is
$e^+e^- \to q\bar q+3g$ (cf.\ Tab.~\ref{tab:colour_dimensions}), for which the 
dimensionality of the basis exceeds the dimensionality of the colour space by one. Thus, we are
able to define a basis with lower dimension by combining basis vectors corresponding to the same
partial amplitude, i.e.,
\begin{equation}\label{eq:reduced_trace}
 \ket{\tilde{b}_\alpha} := \ket{b_{1, 2, \ldots, m}} + (-1)^m \ket{b_{1, m, \ldots, 2}} \equiv \Tr[T_{i_1} T_{i_2} \cdots T_{i_m}] + (-1)^m \Tr[T_{i_1}T_{i_m}\cdots T_{i_2}]\,.
\end{equation}
We refer to this as the \textquotedblleft reduced trace
basis\textquotedblright. In Sec.~\ref{sec:setup_validation}, we report on the
validation of the colour correlators for the specific processes under
investigation. 

\subsubsection*{The multiple-emission function $\mathcal{F}$}
The component $\mathcal{F}$ of Eq.~\eqref{eq:CAESAR} captures the effect of
multiple emissions on the cumulative distribution $\Sigma$. Recursive infrared
collinear safety of the observable guarantees that one can treat emissions below a cutoff
$\epsilon$ as unresolved, such that the cancellation between real and virtual
corrections is complete. To eliminate subleading terms, one would
usually have to find suitable integral transforms to factorise the contributions
of multiple individual emissions. This is for example straightforward for
additive observables, $V(k_1,\dots,k_m)=\sum_{i=1}^m V(k_i)$, where one can insert a
simple integral representation of the corresponding $\Theta$-function. In
general, this procedure can however yield rather untractable expressions, or
there are no such transformations known as is the case for the jet-resolution
scales. In those cases, one can resort to numerical evaluations, rescaling
momenta to eliminate contributions that vanish in the strict soft limit. In
\cite{Banfi:2004yd} a general form of the multiple-emission function
$\mathcal{F}$ suitable for numerical evaluation was derived.  It was in
particular used to resum $y_{23}$ to NLL~\cite{Banfi:2001bz} and (with suitable
additions) NNLL in~\cite{Banfi:2016zlc} accuracy. The general formula for a
specific flavour channel $\delta$ reads 
\begin{equation}
  \begin{split}
    \mathcal{F}^{\mathcal{B}_\delta}(L)=
    \lim_{\epsilon\rightarrow0}
    e^{R'\,\ln\epsilon}\sum_{m=0}^\infty\frac{R^{\prime m}}{m!}\left(
      \prod_{i=1}^{m+1} \sum_{l_i\in \delta} \frac{R^\prime_{l_i}}{R^\prime}\right.&\left. \int_\epsilon^1\frac{\mathop{d\zeta_i}}{\zeta_i}\int_0^1\frac{\mathop{d\xi_i}}{\mathcal{N}_{l_i}}\int_0^{2\pi}\frac{\mathop{d\phi_i}}{2\pi} \right)\delta(\text{ln}\zeta_1)\\
    &\times
    \text{exp}\left(-R'\,\ln\lim_{\bar{v}\rightarrow0}\frac{V(\mathcal{B}_\delta;\kappa_1(\zeta_1\bar{v}),...,\kappa_{m+1}(\zeta_{m+1}\bar{v}))}{\bar{v}}\right),\label{eq:Ff}
  \end{split}
\end{equation}
\begin{equation*}
  \text{with} \hspace{1cm} \mathcal{N}_{l_i} = \left(1+\frac{a+(1-\xi_i)b_{l_i}}{a(a+b_{l_i})}2\lambda\right) \int_0^1
  \text{d}\tilde\xi_i\frac{1}{1+\frac{a+(1-\tilde\xi_i)b_{l_i}}{a(a+b_{l_i})}2\lambda},
\end{equation*}
where $\kappa_i(\zeta_i\bar{v})$ defines an emission from the Born configuration
$\mathcal{B}$ with, according to Eq.~(\ref{eq:V(k)}), an individual contribution to
the observable of $V(\kappa_i(\zeta_i\bar{v}))=\zeta_i\bar{v}$.
The variable $\xi_i$ denotes the fraction of the maximal
rapidity for leg $l_i$. The requirement
$V(\kappa_i(\zeta_i\bar{v}))=\zeta_i\bar{v}$ implies
$\eta_{i,\mathrm{max}}=\ln(1/\zeta_i\bar{v})/(a+b_{l_i})$. At NLL accuracy, a
uniform rapidity boundary $\eta_{i,\mathrm{max}}=\ln(1/\bar{v})/(a+b_{l_i})$
might be assumed for all emissions.\footnote{This ambiguity is one of the
  subleading contributions that vanish in the limit $\bar{v}\to0$. If one
  wants to use the specialised algorithm presented in Ref.~\cite{Banfi:2001bz}, it is
  mandatory to explicitly remove these subleading contributions by
  implementing the same rapidity boundary for all emissions. As we take the
  $\bar{v}\to0$ limit numerically and evaluate the observable exactly, we
  expect to obtain identical (within statistics) functions
  $\mathcal{F}(R^\prime)$ no matter which rapidity boundary we use. We have
  explicitly verified this for the observables considered here. Our tests of
  subleading contributions in Sec.~\ref{sec:setup_validation} are obtained including these subleading contributions,
  i.e.\ $\eta_{i}=\xi_i\ln(1/\zeta_i\bar{v})/(a+b_{l_i})$, consistent with
  e.g. Eqs.~(3.9)-(3.10) of Ref.~\cite{Banfi:2004yd}.}
See also App. \ref{app:formulas} for details on our notation.  
Note  that the normalisation of the $\xi$ integral is trivial if the observable
does not scale with the rapidity in the $l_i$ collinear limit, $b_{l_i}=0
\Rightarrow\mathcal{N}_{l_i}=1$. The channel $\delta$ defines the flavours of
the Born partons and hence their corresponding Casimirs $C_l$. In general,
$\mathcal{F}$ depends on these Casimirs and $L$ separately.
For most obervables however, it turns out that the dependence is only on  the
combination $R^\prime$. We hence also use the notation $\mathcal{F}(R^\prime)$
where no ambiguities can arise. If only the normalisation $d_l$ but not the
actual scaling behaviour depend on $\mathcal{B}$, the multiple-emission function
can be evaluated for every Casimir permutation of a reference Born configuration
$\mathcal{B}_\mathrm{ref}$ and does not have to be calculated on the fly.  This
is generally desirable for this approach to be useful, as avoiding subleading
contributions in $\mathcal{F}$ requires the numerical evaluation of
the limits $\bar{v}\to0$, $\epsilon\to0$ to high accuracy, usually beyond the
limits of double-precision.
We have implemented Eq.~\eqref{eq:Ff} in a Monte Carlo code, independent of the
original \Caesar work, and use this to calculate the 
$\mathcal{F}$ function in our framework. We evaluate the limits  of the
multiple-emission function for a reference Born configuration numerically,
making use of multiple-precision arithmetic, and tabulate $\mathcal{F}$ for a
grid of $R^\prime$ values. This grid is interpolated using cubic Hermite
splines, making additional use of the known monotonicity of $\mathcal{F}$
\cite{Fritsch:1980, Fritsch:1984}.
We have convinced ourselves that our code correctly reproduces
$\mathcal{F}$ functions for simple additive observables with different
parametrisations such as thrust, C- and D-parameter. As a test for non-additive
variables, we reproduced known results for two-jet observables like heavy-hemisphere
mass and thrust-major as well as the three-jet observable thrust-minor. We
further validate the $\mathcal{F}$-functions used here in
Sec.~\ref{sec:setup_validation}. 

\subsection{Aspects of matching to fixed order}\label{sec:matching}

Finally, the matching to fixed-order QCD matrix elements needs to be accomplished in
order to obtain reliable predictions outside logarithmically dominated
phase-space regions, i.e., for hard, non-collinear emissions. In contrast to the
original appproach in \cite{Gerwick:2014gya}, we here aim for
$\mathrm{NLO}+\mathrm{NLL}^\prime$ accuracy and thus resort to matching the
cumulant distribution rather than using a local subtraction-based method. We
briefly discuss possible matching prescriptions widely used in the literature
and introduce the logR scheme that we employ. 

To obtain physical predictions over the full range of the observable, the
endpoint of the resummed prediction $v_\mathrm{max}$ needs to be corrected
to the actual real-emission kinematic endpoint. To this end we modify all
 expressions by subleading contributions (see
e.g.~\cite{Catani:1992ua}). First, for the expansion to vanish at the
endpoint $v_\mathrm{max}$, we need to modify the resummed result to
\begin{equation}
  R^{\mathcal{B}_\delta}(L) \to
  R^{\mathcal{B}_\delta}(L) - \frac{\alphaS}{2\pi}L\left(\frac{v}{v_\mathrm{max}}\right)^p
  \left[\sum_{l\in\delta} C_l \left(\frac{4 B_l}{a+b_l} +
    \frac{4}{a(a+b_l)}\left(\ln \bar{d}_l -b_l\ln \frac{2 E_l}{\mu_Q}\right)\right) +
    \frac{4}{a}\mathcal{S}^{(1)}\right]\,.
\end{equation} 
In the limit $v\to v_\mathrm{max}$ this subtracts the expansion of $\mathcal{S}$
to first order in $\alphaS/2\pi$, $\mathcal{S}^{(1)} L$, and the part of the
expansion of $R(v)$ that is linear in $L$. In the limit $v\to0$ this addition
vanishes as a power correction, leaving the logarithmic accuracy of the expression 
unaffected. The remaining terms are forced to vanish at the endpoint
$v_\mathrm{max}$ by modifying all logarithms to
\begin{equation}
  \ln 1/v \to
  \frac{1}{p}\ln\left[\left(\frac{x_v}{v}\right)^p-\left(\frac{x_v}{v_\mathrm{max}}\right)^p+1\right] \equiv L
\end{equation}
such that $L(v=v_\mathrm{max}) = 0$ and $L \to \ln x_v/v \sim \ln 1/v$ when $v
\ll 1$, and subtracting the corresponding change in the leading logarithm. The
parameter $x_v$ represents the, in principle, arbitrary  normalisation of the
observable; $p$ is an additional parameter that can be varied to estimate
contributions from power corrections.   
There is a certain ambiguity in how to choose the default value for $x_v \approx
{\cal{O}}(1)$. We follow the approach in \cite{Banfi:2004nk} to cancel the $d_l$ dependence of the
resummation formula. In the simplest form, and sufficient for the analysis
presented here, this amounts to $\ln x_v = 1/n \sum_{l\in \delta} \ln \bar{d}_{l}$,
where
\begin{equation}
  \ln \bar{d}_l= \ln d_l + \int\limits_{0}^{2\pi}\frac{d\phi_l}{2\pi}\ln g_l(\phi_l)\,.
\end{equation}

There are different approaches to write matched expressions at $\mathrm{NLO+NLL}^\prime$ for the
cumulative distribution $\Sigma$, i.e., expressions that reproduce the fixed-order
result including terms of order $\alphaS^2$ relative to the respective
Born process and reduce to the NLL result in the limit $v\to0$.
At leading order, we might use a simple additive matching. To express more
involved matching schemes, we introduce the following notation:
$\Sigma^\delta_\mathrm{res,fo,match}$ denotes
the cumulative distribution in resummation, 
at fixed order, or matched between the two, for the (family of) channel(s) $\delta$ as
defined by a suitable jet algorithm. In the following, labels are omitted in general
expressions and we use the shorthand $\sigma = \Sigma(1)$. We denote the expansion of any
$\Sigma$ to order $\alphaS^2$ relative to the $n$-parton Born process as
\begin{equation}
  \Sigma = \Sigma^{(0)} + \Sigma^{(1)} + \Sigma^{(2)} + {\cal{O}}(\alphaS^{n-2+3})\,,
  \hspace{1cm} \Sigma^{(k)} \propto \alphaS^{n-2+k}\,.
\end{equation}
Practically, at least for $\Sigma^{(2)}_\mathrm{fo}$, we only calculate
\begin{equation}
  \overline{\Sigma}^{(2)}_\mathrm{fo} = \int_v^1 \mathop{d\sigma^{(2)}}~, \hspace{1cm}   \Sigma^{(2)}_\mathrm{fo} = \sigma^{(2)}_\mathrm{fo} - \overline{\Sigma}^{(2)}_\mathrm{fo}\,.
\end{equation}
We first define the multiplicative matching scheme,
\begin{equation}
  \Sigma^{\delta}_\mathrm{mult} = \Sigma^{\delta}_\mathrm{res}\left[1 + \frac{(\Sigma^{\delta,(1)}_\mathrm{fo}-\Sigma^{\delta,(1)}_\mathrm{res})}{\sigma^{\delta,(0)}}
     + \frac{1}{\sigma^{\delta,(0)}}
    \left(-\overline{\Sigma}^{\delta,(2)}_\mathrm{fo}-\Sigma^{\delta,(2)}_\mathrm{res} -
    \frac{\Sigma^{\delta,(1)}_\mathrm{res}}{\sigma^{\delta,(0)}} \left(\Sigma^{\delta,(1)}_\mathrm{fo}-\Sigma^{\delta,(1)}_\mathrm{res}\right)\right)\right]\,.
\end{equation}
To order $\alphaS^{(n-2)}\alphaS^2$, we obviously recover the NLO cumulative distribution,
apart from a missing additive constant $\sigma^{\delta,(2)}_\mathrm{fo}$. To the order of our calculations, it is, however, not affecting
the normalised differential distributions we are interested in. In the limit $v\to 0$, it reduces to 
\begin{equation}
  \Sigma_\mathrm{mult}^\delta \to \left(1+\frac{\alphaS}{2\pi}C_1^\delta
  + {\cal{O}}(\alphaS^2)\right)\Sigma_\mathrm{res}^\delta~,\hspace{1cm}  \frac{\alphaS}{2\pi}C_1^\delta \equiv \lim_{v\to 0} \frac{\Sigma^{\delta,(1)}_\mathrm{fo}-\Sigma^{\delta,(1)}_\mathrm{res}}{\sigma^{\delta,(0)}}\,.
\end{equation}
With this scheme, terms of the order $\alphaS^{(n-2)}\alphaS^{k}L^{2k-2}$ are
reproduced 
correctly in the expansion. Note that this relies on the fact that the
leading-logarithmic terms $\propto\alphaS^{(n-2)}\alphaS^{k}L^{2k}$ do not depend on the
kinematics or colour structure of the Born event but only on the flavour
assignment. We are only interested in the terms where
$\frac{\alphaS}{2\pi}C_1^\delta$ multiplies one of those leading logarithms,
all other cross terms are of subleading orders that are not computed
consistently anyway. Thus we do not need to worry about the dependence of
$\frac{\alphaS}{2\pi}C_1^\delta$ on details of the Born event, apart from the
channel $\delta$. See also the discussion in \cite{Banfi:2010xy}.
We refer to this achieved accuracy in the expansion together with the presence of all terms of the form
$\alphaS^kL^{k+1}$ and $\alphaS^{k}L^k$ in $\ln (\Sigma_\mathrm{res})$ as
$\mathrm{NLL}^\prime$ accuracy.

We also define a matching scheme based on the logarithm of the cumulative
distributions, for consistency with the existing literature called LogR matching
scheme, see e.g. \cite{Catani:1992ua,Banfi:2010xy},
\begin{equation}
  \Sigma^{\delta}_\mathrm{LogR} = \Sigma^{\delta}_\mathrm{res}\exp\left(\frac{\Sigma^{\delta,(1)}_\mathrm{fo}-\Sigma^{\delta,(1)}_\mathrm{res}}{\sigma^{\delta,(0)}}
    \right) \exp\left[\frac{-\overline{\Sigma}^{\delta,(2)}_\mathrm{fo}-\Sigma^{\delta,(2)}_\mathrm{res} -
    \frac{\left(\Sigma^{\delta,(1)}_\mathrm{fo}\right)^2-\left(\Sigma^{\delta,(1)}_\mathrm{res}\right)^2}{2\sigma^{\delta,(0)}} }{\sigma^{\delta,(0)}}
    \right]\,.
\end{equation}
This defines our default matching choice in the evaluation of
jet-resolution scales. The arguments on the achieved accuracy apply as for the multiplicative
matching after expanding. The final distributions are then obtained by adding the different channels
\begin{equation}\label{eq:match_final}
  \Sigma_\mathrm{match} = \sum_{\delta\in\mathcal{B}} \Sigma_\mathrm{match}^{\delta} + \sum_{\delta\notin\mathcal{B}} \Sigma^{\delta}_\mathrm{fo}\,,
\end{equation}
where the second sum accounts for channels not corresponding to any
structure found in the physical Born process, but permitted by the jet-clustering algorithm.
We will later on present normalised differential distributions given by
\begin{equation}
  \frac{1}{\sigma}\frac{\mathop{d\sigma}}{\mathop{d\ln v}} = \frac{1}{\Sigma_\mathrm{match}(1)}
  \frac{\mathop{d\Sigma_\mathrm{match}(v)}}{\mathop{d\ln v}}\,.
\end{equation}

\section{Resummation of Durham jet-resolution scales}\label{sec:resummed_results}

In this section we want to specify the general \Caesar formalism for the
case of Durham jet-resolution scales. In particular we discuss the actual
observables and introduce our choices for parameters and scales. This is
supplemented by the validation of the soft-colour correlators and the
$\mathcal{F}$ functions we employ in the resummation of $y_{34}$, $y_{45}$ and
$y_{56}$.

\subsection{Definition of the observables}\label{sec:definition_observable}

The functional form of the jet-resolution scales we consider here has been given
in Eq.~\eqref{eq:jet_def}. We restrict ourselves to what is commonly referred to
as the E-scheme,  i.e., jets are combined by adding their four-momenta. To
completely define the observable $y_{n,n+1}$, we first require $y_{n-1,n} >
y_\mathrm{cut}$ to define a hard $n$-parton configuration around which the resummation is
performed. From the point of view of an experimental measurement, it would be
tempting to choose a rather small value of $y_\mathrm{cut}$ to get a sizable
cross section. However, besides the obvious need to regularise the infrared
divergences in the Born event, cf.\ the jet-function $\mathcal{H}$ in
Eq.~\eqref{eq:CAESAR}, the formalism we use here is based on the assumption that
additional radiation is soft relative to all legs. In particular, configurations
with $y_{n,n+1}\approx y_{n-1,n}$ need to be described by the achieved
fixed-order accuracy, and configurations where the production of one of the Born
jets is already logarithmically enhanced need to be avoided. This prompts for
larger values of $y_\mathrm{cut}$. In practice, for our theoretical studies, we
find an observable definition with $y_\mathrm{cut}=0.02$ a good compromise
between these considerations, 
although admittedly smaller cuts would have to be investigated for a realistic
experimental analysis, in particular for the highest multiplicities. Starting
from this type of events, we are then interested in the next hardest resolution
scale. If this $(n+1)$th jet consists of only one gluon that is soft and/or
collinear to one of the legs $l=1\dots n$, one indeed obtains the general form
of observables in the \Caesar formalism in Eq.~\eqref{eq:CAESAR}, with the
parameters $g_l=1, b_l=0$ and $a=2$, i.e.,
\begin{equation}\label{eq:scaling}
 V(k) \equiv y_{n,n+1} = \frac{k_T^2}{Q^2}\,.
\end{equation}
In this case, all parameters are independent of the leg $l$ relative to
which the emission is soft and/or collinear. Note that in Eq.~\eqref{eq:scaling},
we have explicitly normalised by the squared centre-of-mass energy
$Q^2$. This corresponds to fixing the coefficients $d_l$ in Eq.~\eqref{eq:V(k)} to
$d_l = \mu_Q^2/Q^2$. The explicit form of the radiator $R_l$ for an
observable with the scaling behaviour given in Eq.~\eqref{eq:scaling}, is
specified in App.~\ref{app:formulas} using our conventions.

\subsection{Calculational setup, parameter and scale choices}\label{sec:setup_validation}

For $2$-jet observables in $e^+e^-$ annihilation, there is almost no ambiguity in the
choice of scales, with the centre-of-mass energy $Q=\sqrt{s}$ basically being the only
physical scale present in the Born process. For the multijet processes we consider here,
this is no longer valid; due to phase-space constraints, some of the jets
have to be associated with significantly lower scales.

Our default choice for the resummation scale for the variable $y_{n,n+1}$ is
$\mu_Q^2 =y_{n-1,n}Q^2$. As $d_l$ is the same for all $l$, i.e., $d_l\equiv d$, we
have $x_v = d = y_{n-1,n}$. This means that the logarithms are effectively of the form
$\ln\left(y_{n-1,n}/y_{n,n+1}\right)$. Our considered Born processes feature up to
five jet thus constitute a severe multi-scale problem. We address this by choosing
the renormalisation scale $\mu_R^2$ according to the CKKW prescription~\cite{Catani:2001cc},
i.e., based on the nodal Durham jet-resolutions of the $n$-parton Born process. For
the observable $y_{n,n+1}$, this results in
\begin{equation}
  \alphaS(\mu_R^2)^{(n-2)} = \alphaS(y_{23}Q^2)\cdot \ldots \cdot\alphaS(y_{n-1,n}Q^2)\,,
\end{equation}
which is solved for $\mu_R^2$ assuming leading-order running of $\alphaS$, thus
\begin{equation}
  \mu_{R}^2 = \Lambda_\mathrm{QCD}^2 \exp\left[\frac{\prod_{i=3}^n (1-\lambda_i)^{1/(n-2)}}{\alphaS\beta_0}\right]\,,
\end{equation}
with $\lambda_i = \alphaS\beta_0\ln y_{i-1,i}$ and $\Lambda_\mathrm{QCD}^2 =
Q^2 \exp(-1/\alphaS\beta_0)$. The strong coupling $\alphaS$ is evolved at two-loop
accuracy with a fixed number of $\nf=5$ massless flavours.

We fix the endpoint of the resummed distribution to
$y_\mathrm{max}=\min(y_\mathrm{kin},y_{n-1,n})$, where $y_\mathrm{kin}$ denotes
the maximal value of $y_{n,n+1}$ with equal energies for all legs, $E =
Q/(n+1)$. The second constraint enforces $y_{n,n+1} < y_{n-1,n}$. Note, this condition is automatically satisfied for our central scale choices.

To identify the channels $\delta$ in the fixed-order calculations, we use the
jet algorithm described in \cite{Banfi:2006hf} for $e^+e^-$ collisions. In our
default setup, we restrict the jet algorithm to produce jets with at most one
flavour, and additionally require that the flavour assignment is compatible with
a $Z$ decay, i.e., that there is at least one pair of jets identified as quark
and antiquark jets with the same flavour. With these requirements, the second
sum in Eq. \eqref{eq:match_final} vanishes. As we treat all active quarks as
massless, we can ignore the different flavours and only need to identify jets as
either quark- or gluon-like. As there is also no dependence on the relative
energy ordering of the legs, see in particular the discussion on the
$\mathcal{F}$ function, we can collect all events with the same number of quarks
and gluons into one family of channels $\delta$.

The calculation of $\Sigma_\mathrm{res}$ proceeds as described in the previous
section. The fixed-order calculation is also done within the \Sherpa framework,
making use of the Catani--Seymour subtraction scheme~\cite{Catani:1996vz} as
implemented in \Comix~\cite{Gleisberg:2008fv}. We use \OpenLoops~\cite{Buccioni:2019sur}
for one-loop virtual corrections to 3-, 4-, and 5-parton matrix elements. Virtual
corrections to 6-parton matrix elements are generated with \Recola~\cite{Actis:2012qn, Actis:2016mpe,Biedermann:2017yoi}. 
Note that $\sigma^{(2)}$ is not needed in our matching formulas, and hence there
is no need to compute any purely virtual corrections of order
$\alphaS^{(n-2)}\alphaS^2$,
while still achieving NLO accuracy in the normalised differential distribution. 

\subsubsection*{Validation of soft-colour correlators}
\begin{figure}[ht]%
	\centering
	\includegraphics[width=0.45\textwidth]{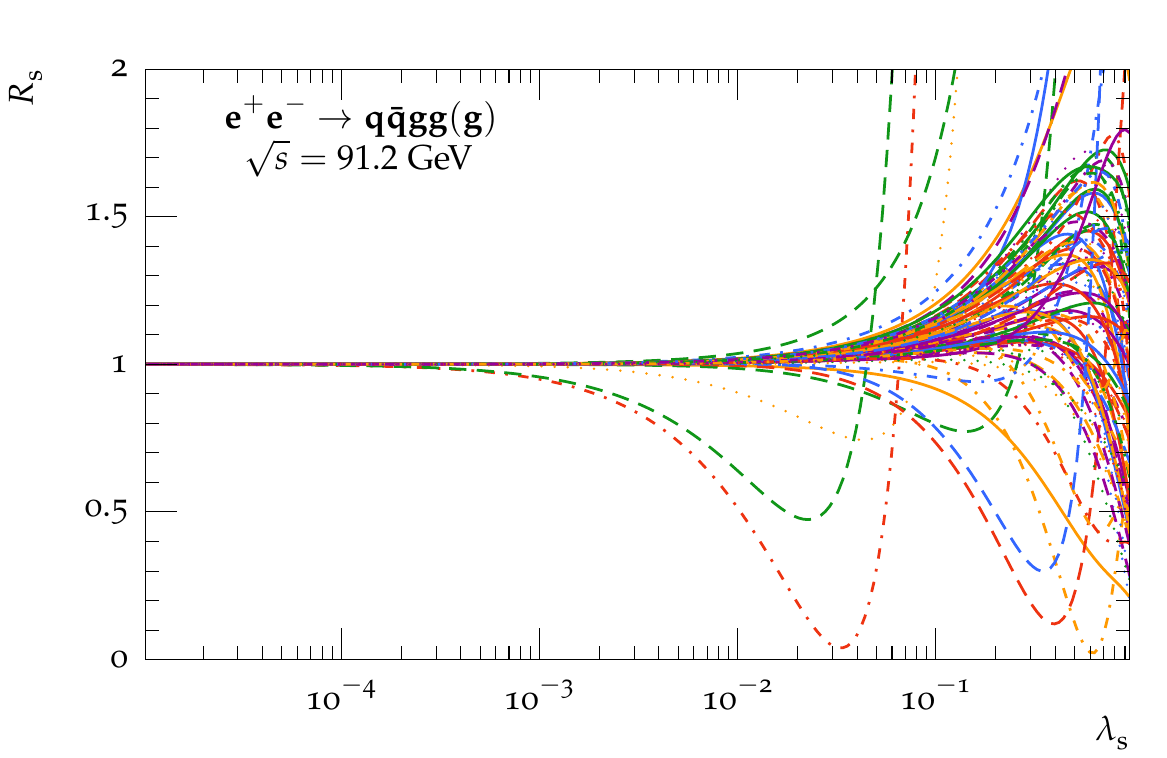}
	\includegraphics[width=0.45\textwidth]{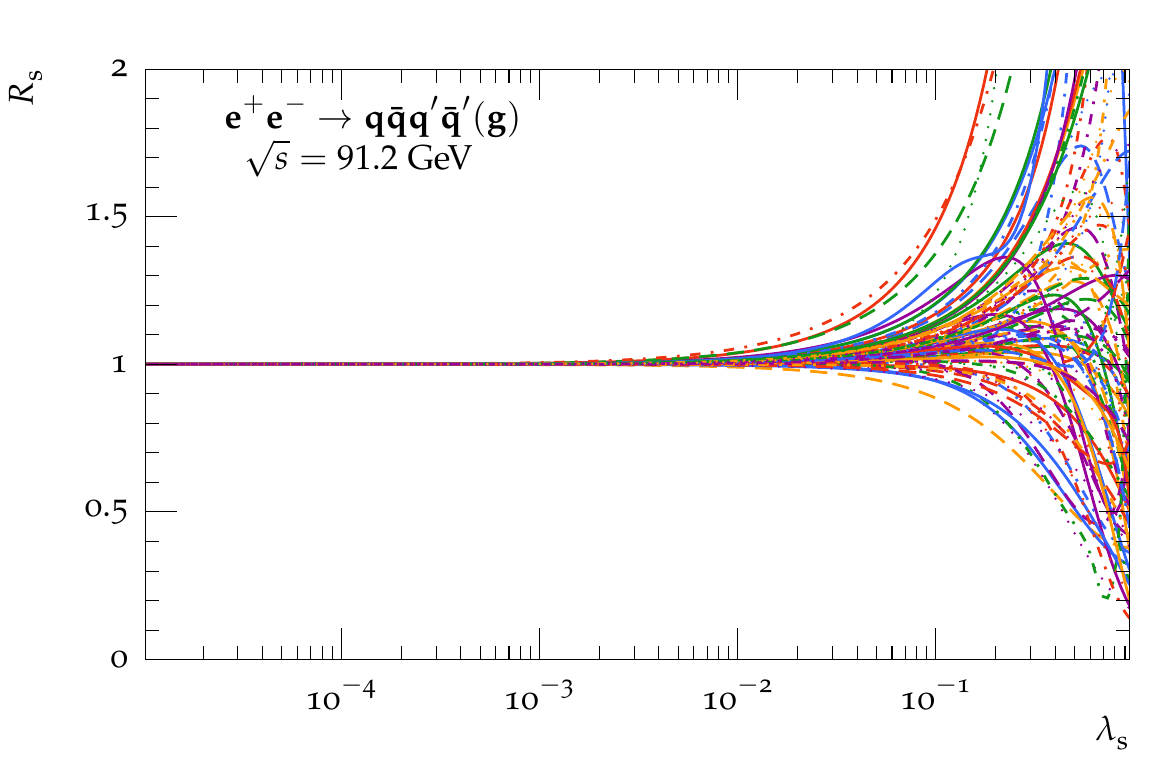}
	\\
	\includegraphics[width=0.45\textwidth]{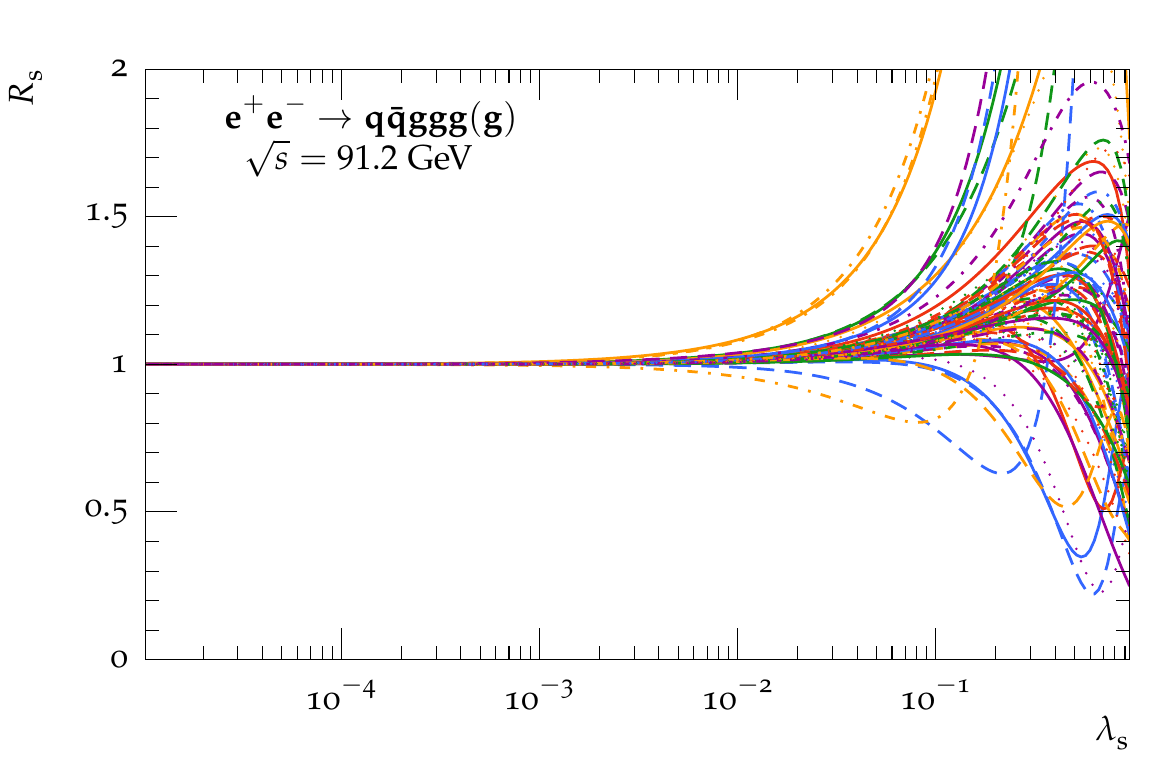}
	\includegraphics[width=0.45\textwidth]{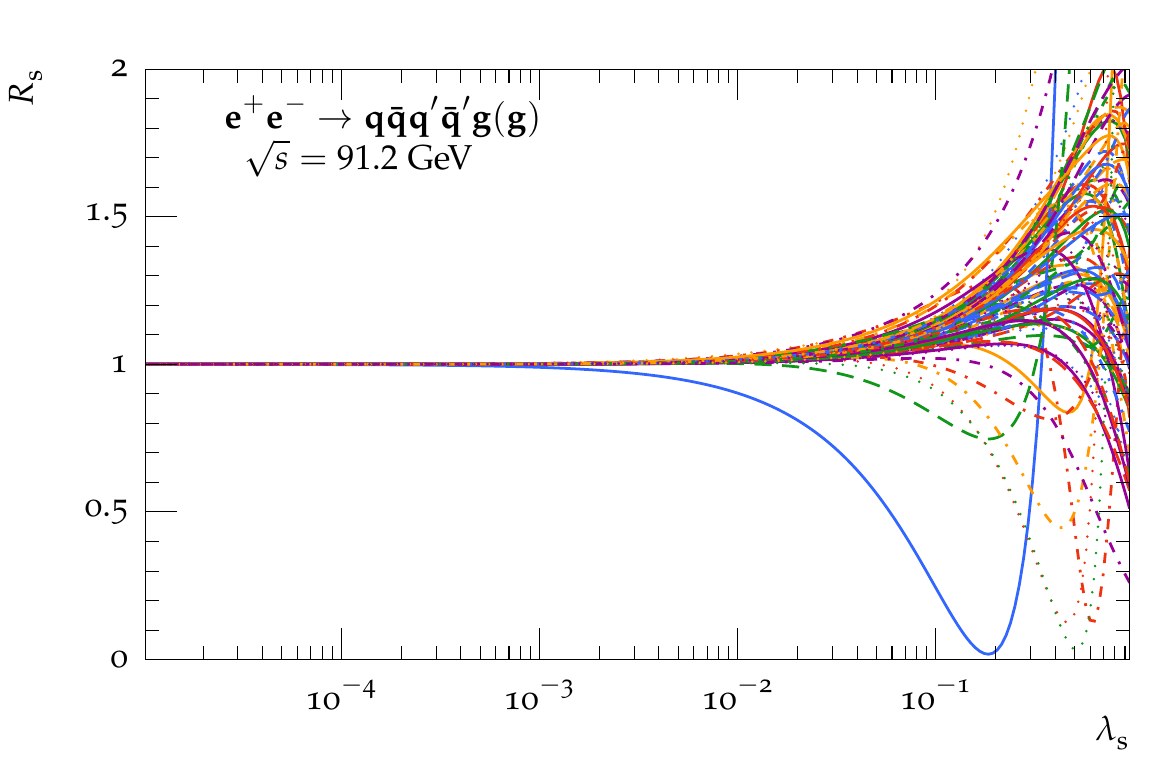}
	\caption{Ratio of the eikonal approximation to the exact $e^+e^- \to 5-,6-\mathrm{parton}$ tree-level
		matrix element versus the softness parameter $\lambdas$.}%
	\label{fig:rratios_durham}%
\end{figure}

We validate all non-trivial colour correlators in Eq.~\eqref{eq:soft_gamma} by comparing
the eikonal approximation to exact $(n+1)$ tree-level matrix elements (see also Ref. \cite{Gerwick:2014gya}),
in form of the ratio
\begin{equation}
\Rs = \frac{\Tr[\gamma c_n H_n]}{\Tr[c_{n+1} H_{n+1}]}\,, \label{eq:RsRatio}
\end{equation}
with the squared eikonal current
\begin{equation}
\gamma = -2\gS^2 \sum\limits_{\substack{i<j}} \frac{p_i \cdot p_j}{(p_i \cdot k_s)(p_j \cdot k_s)} \sum \limits_{\alpha, \beta}\bra{b_\alpha}\boldsymbol T_i \boldsymbol T_j\ket{b_\beta}c^{\alpha\beta}\,. \label{eq:soft_gamma}
\end{equation}
We randomly pick 100 distinct, non-collinear $(n+1)$-parton configurations, regularised by
$y_{\mathrm{cut}}=0.02$, and scale the momentum $k_s$ of the first gluon in the amplitude by a softness parameter $\lambdas$, i.e., $k_s \mapsto \lambdas k_s$, $\lambdas \to 0$.
This gluon is assumed to be emitted by the dipole spanned by its direct neighbours $s_-$ and $s_+$ which absorb its recoil by
\begin{equation}
\begin{split}
p_{s_-}' &= p_{s_-} - k_s + \frac{p_{s_-} \cdot k_s}{p_{s_+} \cdot (p_{s_-} - k_s)} p_{s_+}\,, \\
p_{s_+}' &= \left(1- \frac{p_{s_-} \cdot k_s}{p_{s_+} \cdot (p_{s_-} - k_s)} \right)p_{s_+}\,.
\end{split} \label{eq:GluonRecoil}
\end{equation}
In the limit of soft, non-collinear kinematics, the ratio in Eq.~(\ref{eq:RsRatio}) has to approach unity by the factorisation theorem of QCD.
This is indeed observed for all the partonic channels relevant for the resummation of $y_{34}$, $y_{45}$ and $y_{56}$ in $e^+e^-$ annihilation.
Fig.~\ref{fig:rratios_durham} contains the validation of all non-trivial colour contributions to partonic configurations relevant for $5$- and $6$-jet production,
respectively. 

\begin{figure}
	\centering
	\includegraphics[width=.49\textwidth]{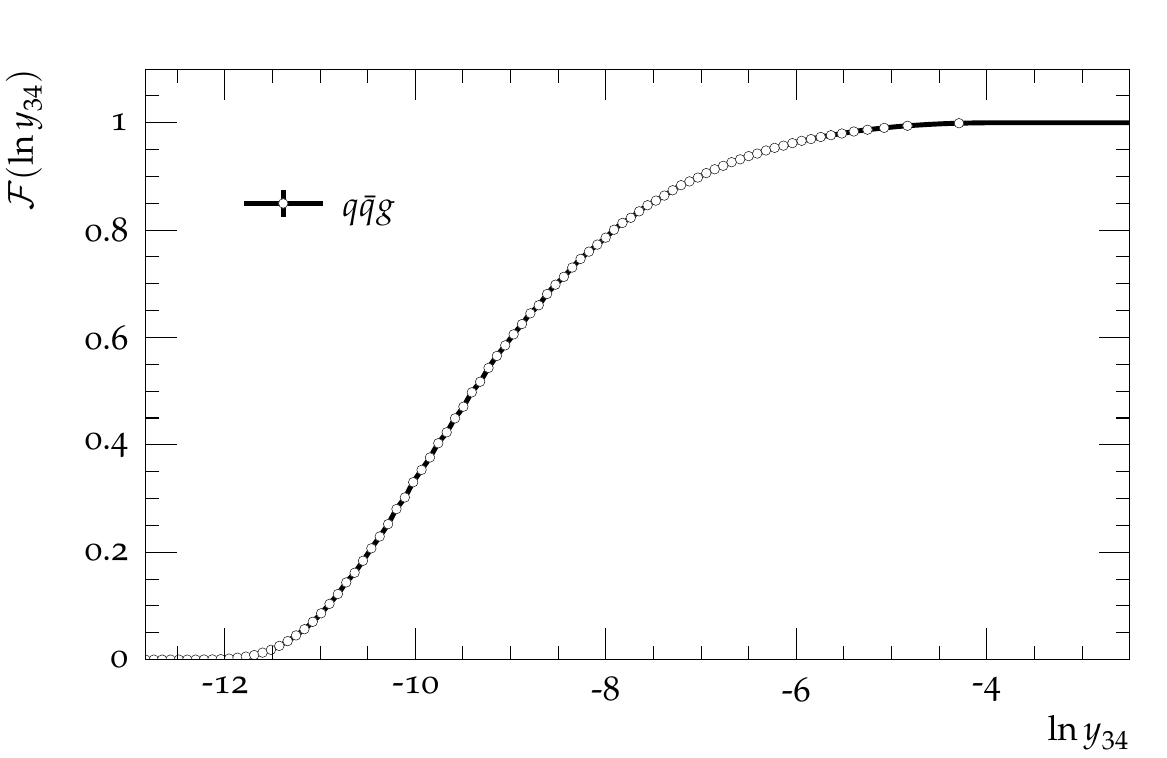}
	\includegraphics[width=.49\textwidth]{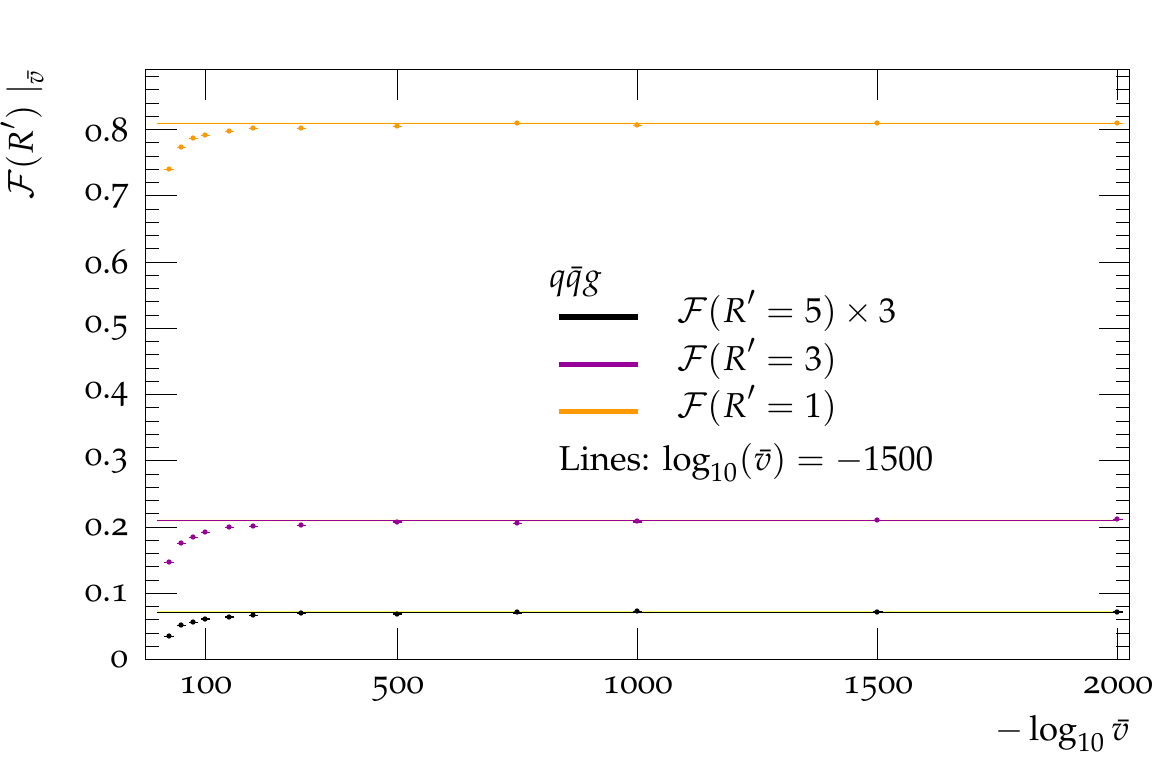}
	\includegraphics[width=.49\textwidth]{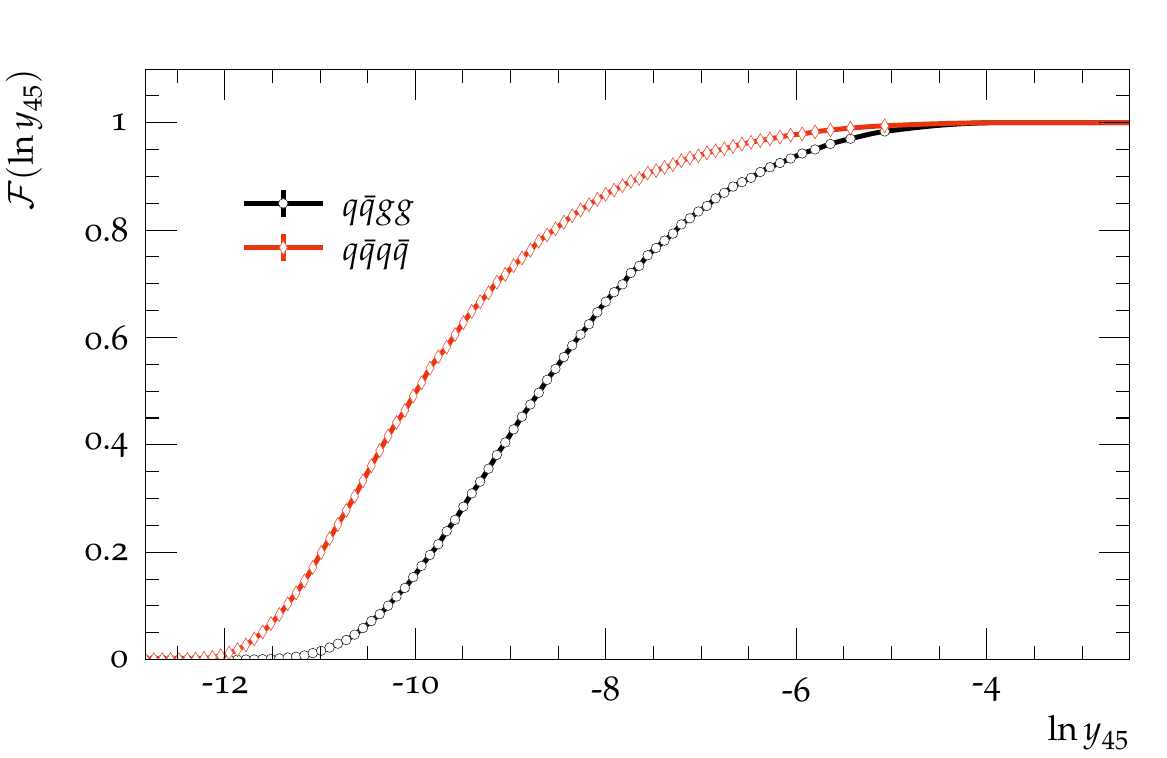}
	\includegraphics[width=.49\textwidth]{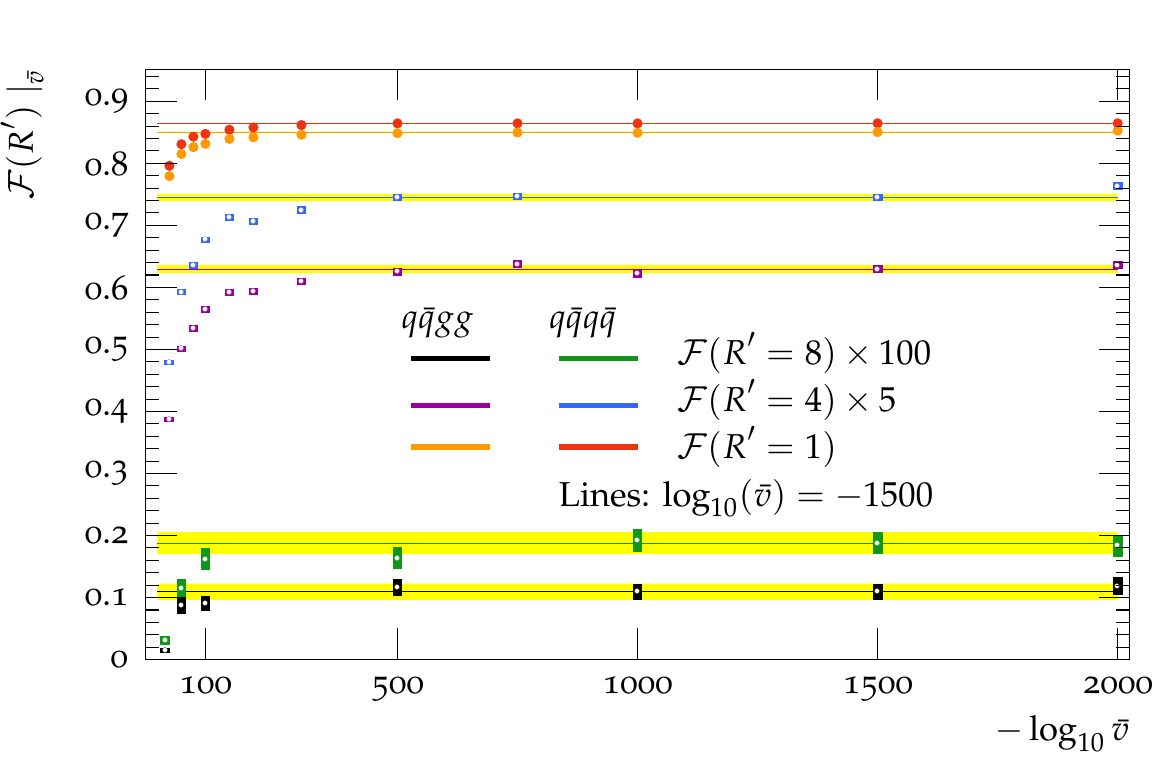}
	\includegraphics[width=.49\textwidth]{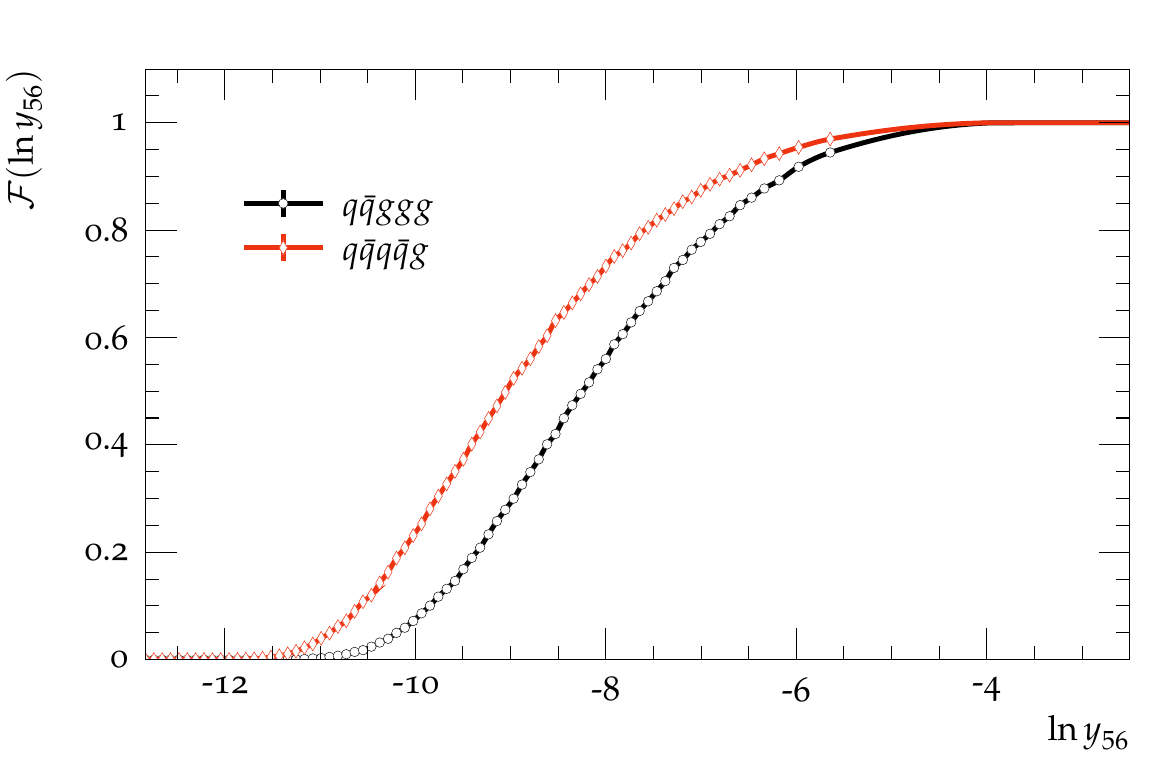}
	\includegraphics[width=.49\textwidth]{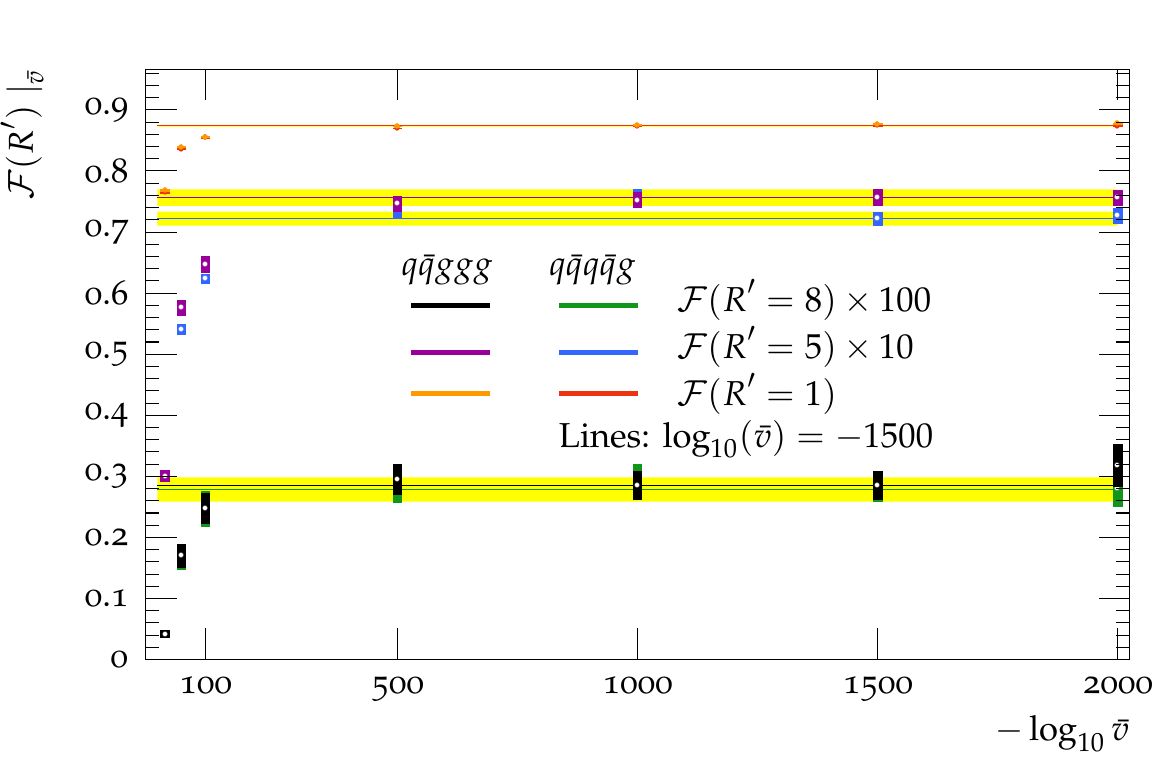}
	\caption{Multiple-emission function $\mathcal{F}$ for the different
		Born channels $\delta$ appearing in the resummation of $y_{34}$, $y_{45}$,
		and $y_{56}$, respectively. While the left column emphasises the dependence
		on the large logarithm, i.e., $\ln(y_{n,n+1})$, the right column shows
		the convergence of the numerical evaluation of the limit $\bar{v}\to 0$ for
		selected values of $R'$. The lines with error bands mark the
                value and statistical uncertainty for $\bar{v}=10^{-1500}$. Note
                that for $y_{34}$ as well as the other $\mathcal{F}$ functions
                with $R^\prime=1$, the dot sizes are larger than the error bar
                would be.}\label{fig:FFunction}
\end{figure}

\subsubsection*{Validation and results for the multiple-emission function}
The $\mathcal{F}$ function is evaluated numerically for all relevant
multiplicities and, where needed, different channels $\delta$.
In the limit $\bar{v}\rightarrow 0$, all emissions become collinear 
to their hard legs and emissions are only clustered together if they originate from
the same parton in the Born event. From this, it is straight-forward to see that
the multiple-emission function is independent of details of the Born kinematics
and can thus be evaluated for a reference configuration for each flavour channel
$\delta$. This amounts to calculate $\mathcal{F}$ for every possible number of
external gluons. Applying the condition of emissions being only clustered if they
were emitted from the same Born parton explicitly removes any dependence on the
used kinematics.

For reference, the numerical results for $\mathcal{F}(L)$ are shown in
Fig.~\ref{fig:FFunction}, for a configuration with $\mu_R^2 = \mu_Q^2 =
y_\mathrm{cut} Q^2$. This could be converted to $\mathcal{F}(R^\prime)$ using
Eq.~\eqref{eq:Rprime}. In the same figure, we also show how $\mathcal{F}(R^\prime)$
approaches the $\bar{v}\to0$ limit for several representative $R^\prime$
values. It is not feasible to stick to double precision such that multiple-precision
arithmetics have to be used. However we are observing
convergence starting from values of $\bar{v}\approx
10^{-500}$. It appears that, when interpreted as a function of $R^\prime$, the
dependence of $\mathcal{F}$ on the Born channel is almost insignificant, in
particular for $y_{56}$. Note, however, that the two functions are not exactly
equal in any case and that this similarity depends on the relatively close numerical
values of the Casimirs $\CF$ and $\CA$.

In the numerical evaluation of Eq. \eqref{eq:Ff}, the limits of the
soft scaling $\bar{v}\to 0$, the cutoff $\epsilon \to 0$, and the number points used
in the Monte Carlo integration $N_\mathrm{points}\to \infty$ are taken numerically,
thus replacing $0$ ($\infty$) with a very small (large) but finite value. Finite
Monte Carlo statistics limit the
numerical accuracy of the integral evaluation, while subleading
contributions are introduced for non-zero $\bar{v}$ and $\epsilon$.
This causes a systematic difference with respect to taking the exact limit
as done for instance for the $y_{23}$ resolution scale in
Ref.~\cite{Banfi:2001bz}. Although, one could imagine implementing the
algorithm used there for higher parton multiplicities to avoid any subleading
contributions, we opt for a fully generic approach here. We argue that subleading
contributions can be controlled to an extent that they become insignificant
in final results. However, given that smaller values of $\bar{v}$ and $\epsilon$ 
increase the computation time per Monte Carlo point, a compromise
between eliminating subleading contributions and the number
of points $N_\mathrm{points}$ we can reasonably evaluate given the
computing resources available needs to be found. 
In the right-hand panel of Fig.~\ref{fig:FFunction}, we demonstrate that,
indeed, for sufficiently small $\bar{v}$ values $\mathcal{F}$ reaches a plateau and
no longer significantly depends on the very value of $\bar{v}$.
The remaining subleading contributions are of order of the statistical
precision used to check the residual dependence of $\mathcal{F}$. 
At the same time, we verify that the remaining statistical uncertainty of
the Monte Carlo evaluation of $\mathcal{F}$ is insignificant for our final
resummed prediction. 

\begin{figure}
  \centering
  \includegraphics[width=.49\textwidth]{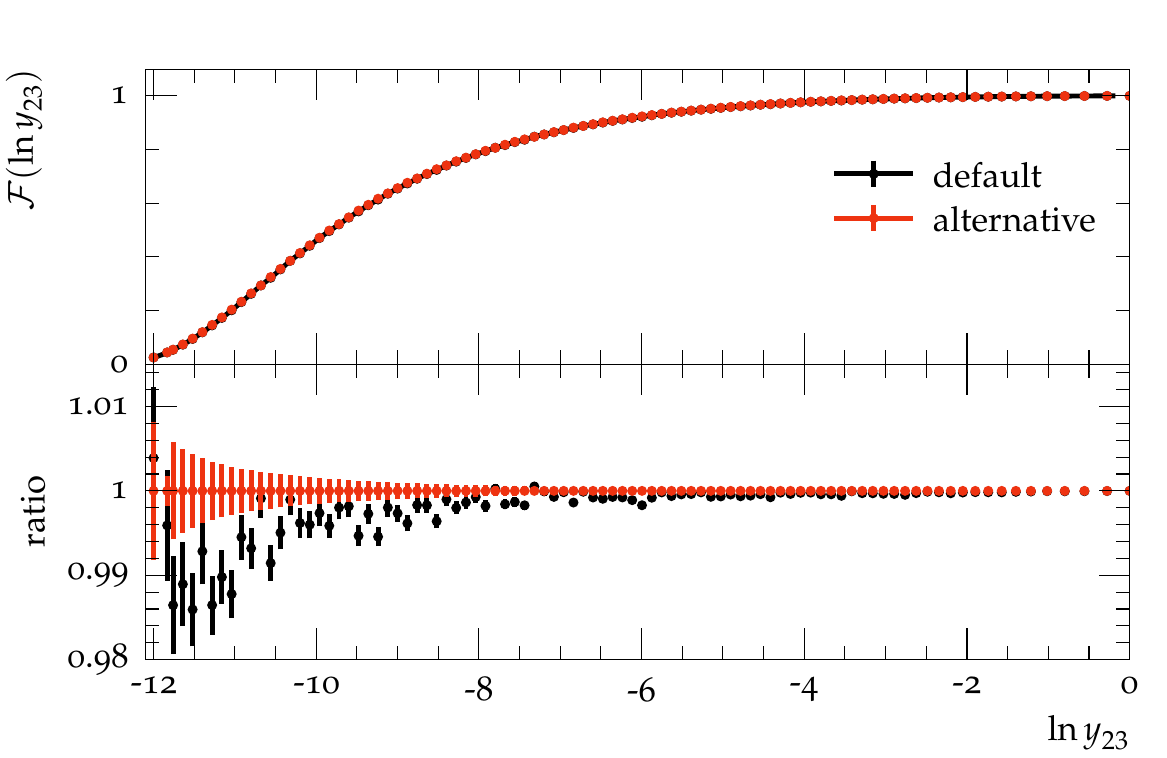}
  \includegraphics[width=.49\textwidth]{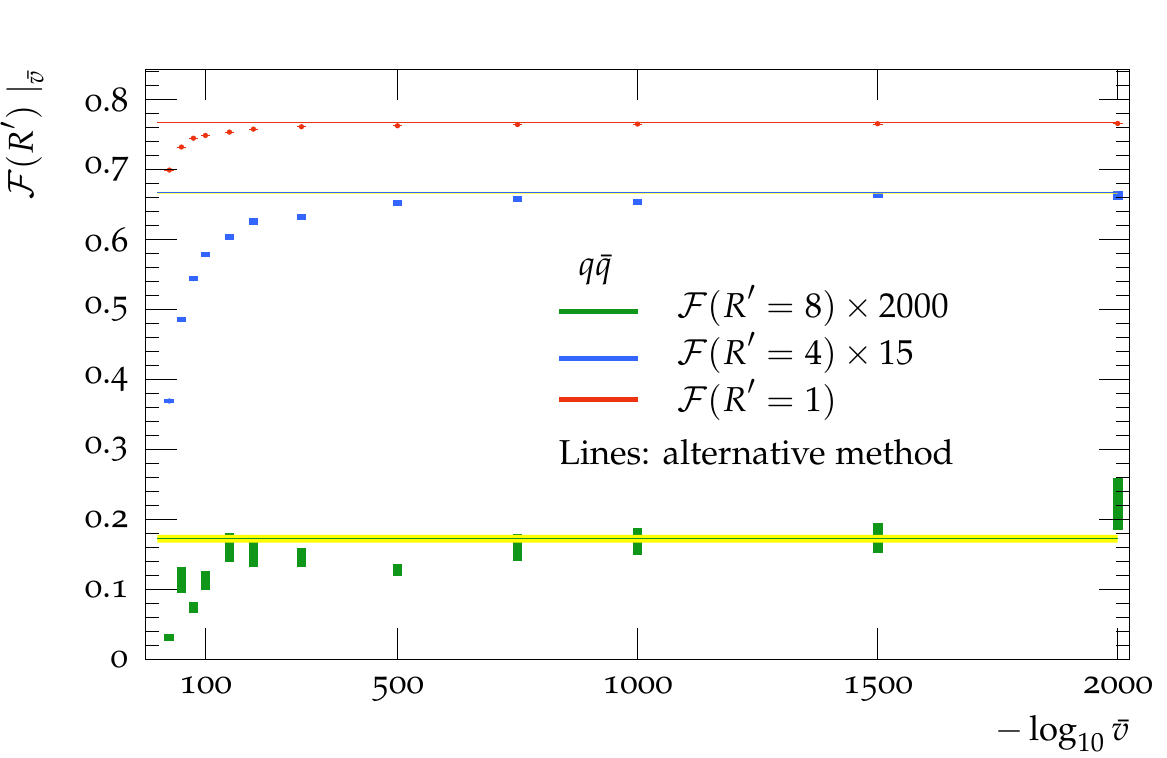}
  \caption{Multiple-emission function $\mathcal{F}$ appearing in the
    resummation of $y_{23}$, evaluated with our default numerical algorithm
    and the alternative method described in \cite{Banfi:2001bz}.
    The left panel presents the dependence on the large logarithm,
    i.e., $\ln(y_{23})$, the right panel shows the convergence of the
    numerical evaluation in the limit $\bar{v}\to 0$ for selected values of
    $R'$. The lines with error bands mark the value and statistical uncertainty
    when using the exact NLL algorithm presented in~\cite{Banfi:2001bz}.}\label{fig:FFunction_y23}
\end{figure}

To validate our procedure, we compare our default approach to evaluate $\mathcal{F}$
with the algorithm described in \cite{Banfi:2001bz} for the specific case of the observable $y_{23}$. 
We use our default set of numerical parameters and compare the two algorithms in the
left panel of Fig.~\ref{fig:FFunction_y23}, observing that they indeed approach the same values. 
The remaining differences are of the same order of magnitude as the statistical uncertainty.
Given our target statistical accuracy for $\mathcal{F}$ of order 1\% subleading terms
here do indeed not exceed the 2\% level.

To match to NLO fixed-order distributions, we also need the expansion of
$\mathcal{F}$. The argument of \cite{Banfi:2001bz} applies to higher
multiplicities, and we can generally write 
\begin{equation}
  \mathcal{F} = 1 + \sum_{p=2}^\infty \mathcal{F}_p R^{\prime p}~, \hspace{1cm} \mathcal{F}_2 = -\frac{\pi^2}{8} \frac{\sum_{l\in\delta} C_l^2}{2\left(\sum_{l\in\delta} C_l\right)^2}\,.
\end{equation}
Numerically, we checked that these values are correctly reproduced in the
corresponding integrals, again validating the limit.

\section{Predictions for Durham jet-resolution scales}\label{sec:results}

In this section, we present our results for the $4-$, $5-$, and $6-$jet Durham resolution scales.
Our central results, the resummed predictions to $\mathrm{NLO+NLL}^\prime$ accuracy, are presented in
Sec.~\ref{sec:resummed_preds}; we study the influence of LO and NLO matching as well as
subleading colour contributions in Sec.~\ref{sec:impact_matching} and Sec.~\ref{sec:impact_colour}, respectively;
in Sec.~\ref{sec:comparison_PS}, we then compare the resummed predictions with parton-shower simulations
from \Sherpa and \Vincia, where we also address the impact of hadronisation corrections on the parton-level
distributions.

\subsection{Resummed predictions for multijet resolutions}\label{sec:resummed_preds}

\begin{figure}[ht]
	\centering
	\includegraphics[width=.49\textwidth]{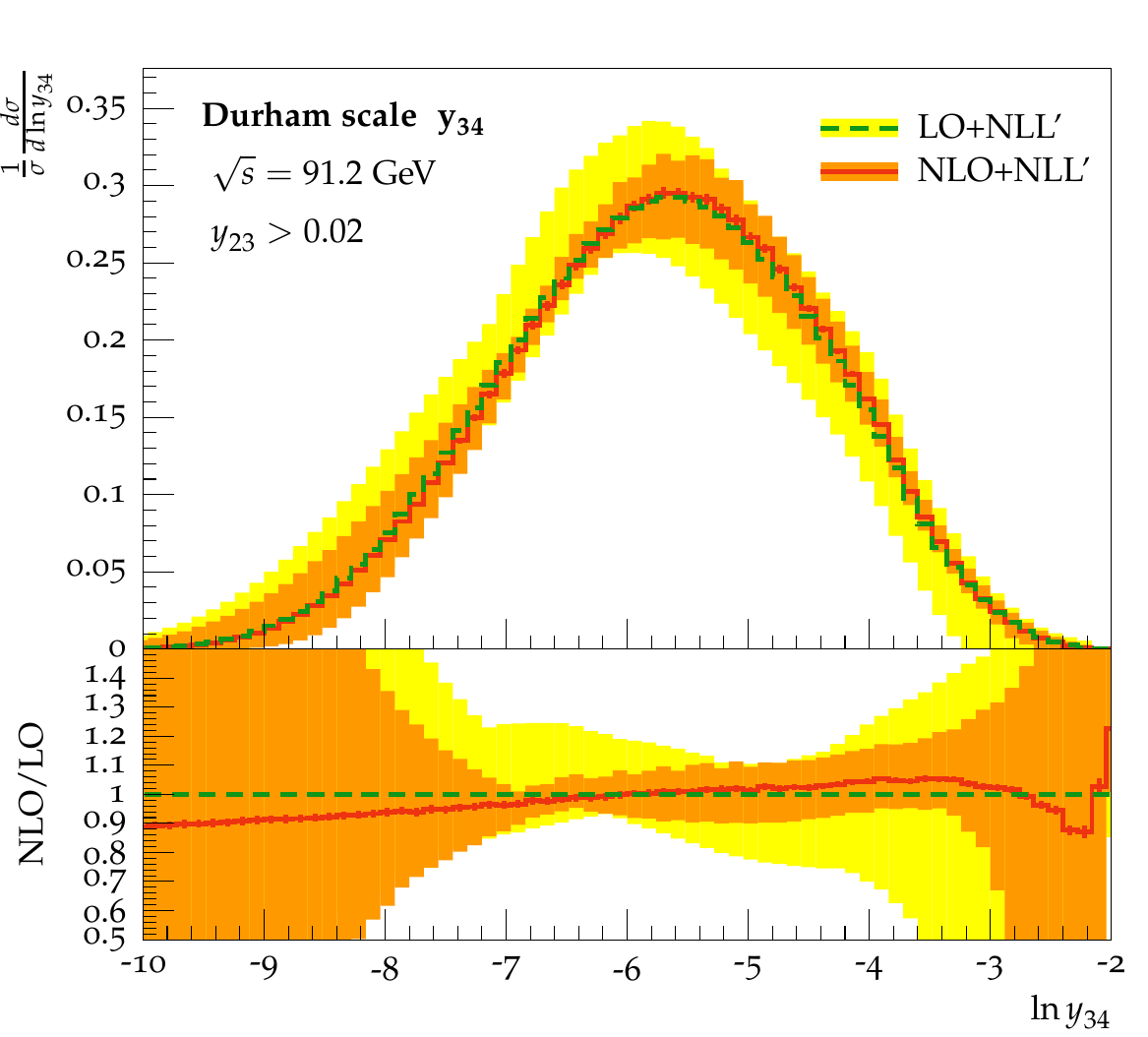}
	\includegraphics[width=.49\textwidth]{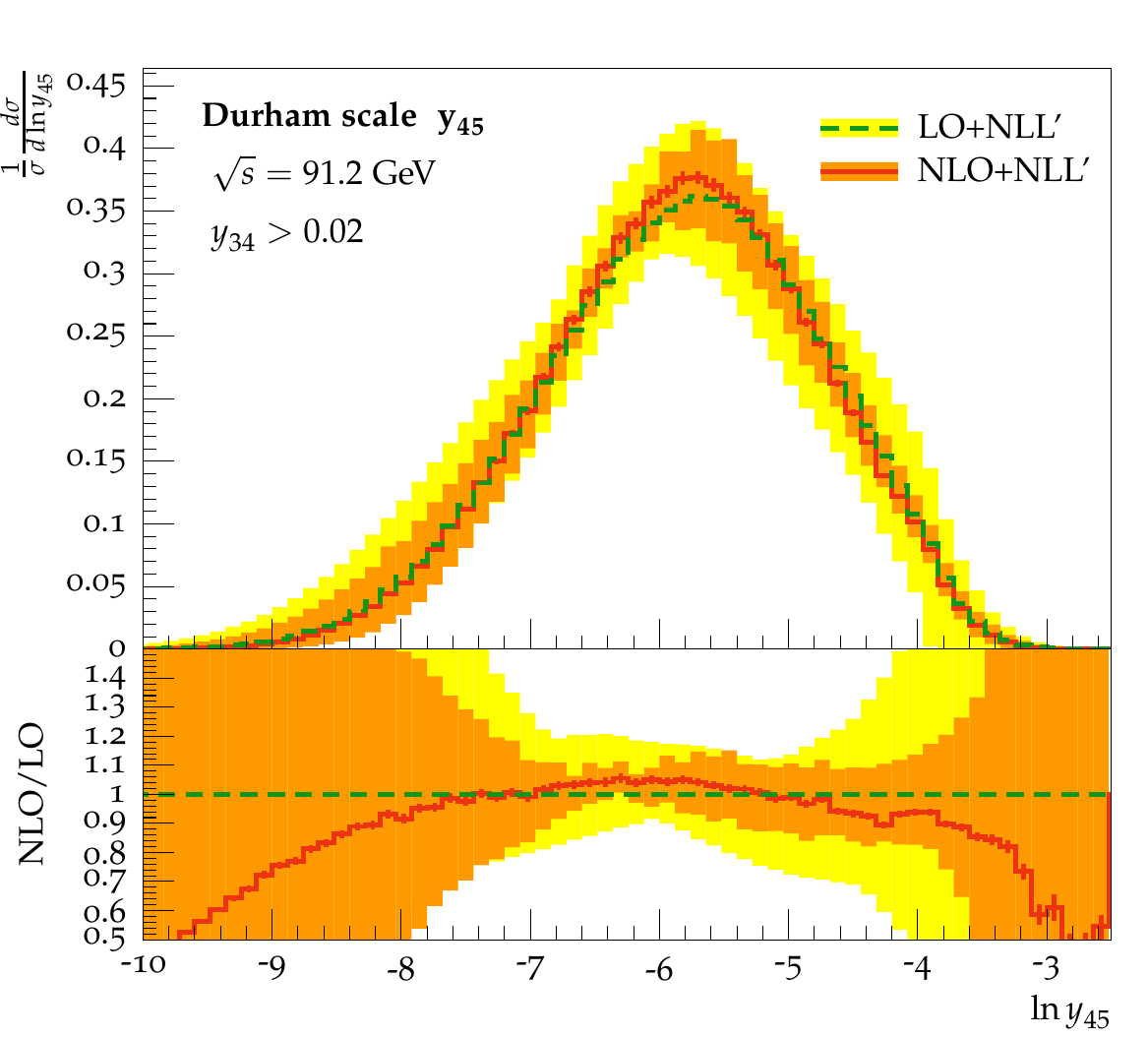}
	\includegraphics[width=.49\textwidth]{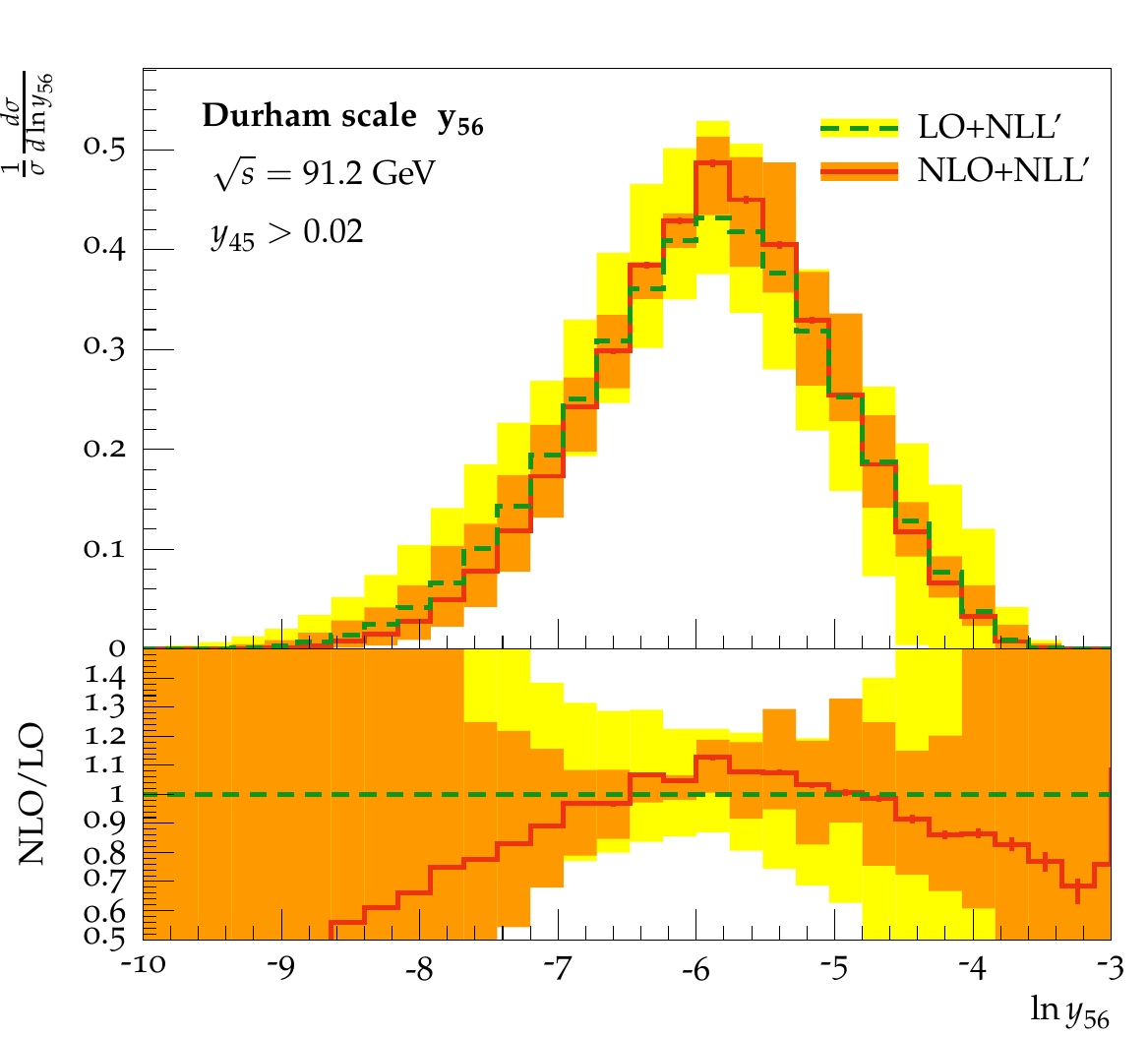}
	\caption{Resummed predictions for the Durham jet-resolutions $y_{34}$,
	  $y_{45}$, and $y_{56}$ in electron-positron annihilation
          at $\sqrt{s}=91.2$ GeV. The lower panels show the local NLO
	  K-factor.}\label{fig:resummed_results}
\end{figure}

Our central results are the resummed distributions of the Durham jet resolutions
$y_{34}$, $y_{45}$, and $y_{56}$, in the definition discussed in
Sec.~\ref{sec:definition_observable}. In particular, we require the Born events
to possess $n=3,4,5$ jets separated by at least $y_\mathrm{cut}=0.02$ in the
calculation for $y_{n,n+1}$, respectively. We perform our calculation at the LEP1
centre-of-mass energy $\sqrt{s}=91.2~\mathrm{GeV}$. We use the LogR matching
scheme as described in Sec.~\ref{sec:matching} to obtain physical distributions
over the full range of the observable.
Fig.~\ref{fig:resummed_results} shows our predictions at LO and NLO, matched to
NLL by the means of the preceding sections to achieve an accuracy of
$\mathrm{LO+NLL}^\prime$ and $\mathrm{NLO+NLL}^\prime$, respectively. To
estimate theoretical uncertainties from missing higher-order corrections, we
consider independent variations of $\mu_R$ , $\mu_Q$, and $p$ by factors of
$0.5$ and $2$, resulting in the yellow ($\mathrm{LO+NLL}^\prime$) and orange
($\mathrm{NLO+NLL}^\prime$) bands.

For higher jet multiplicities, we observe a
narrowing of the distributions as well as a growing impact of NLO contributions
compared to the LO result, being significant only for $y_{45}$ and $y_{56}$. In
the peak region, the impact of NLO corrections stays rather small in any case,
being of order of a few percent only. Regarding $y_{34}$, the NLO corrections 
do not have a large impact on the central prediction, we do, however, observe a
considerable reduction of the theoretical uncertainties, at least away from the very soft
region, where higher-logarithmic corrections become significant. This
expected reduction of uncertainties is observed for the higher multiplicities as well.
Here, the impact of NLO corrections on the matched distributions become larger
when approaching the kinematic endpoint, $y_{n,n+1}\approx y_\mathrm{kin}$, but
consistently stay within the LO uncertainty band.

The remaining statistical uncertainty from the numerical determination of the 
$\mathcal{F}$-function is propagated through to the final result, indicated by the error
bars on the $\mathrm{NLO+NLL}^\prime$ prediction. It is entirely negligible
compared to the overall uncertainty from scale variations. In light of our
earlier discussion of subleading effects emerging from taking the limits in the
evaluation of $\mathcal{F}$ numerically, we conclude that these are indeed
insignificant for our final predictions.

\subsection{Impact of fixed-order corrections}\label{sec:impact_matching}

\begin{figure}[ht]
  \centering
  \includegraphics[width=.49\textwidth]{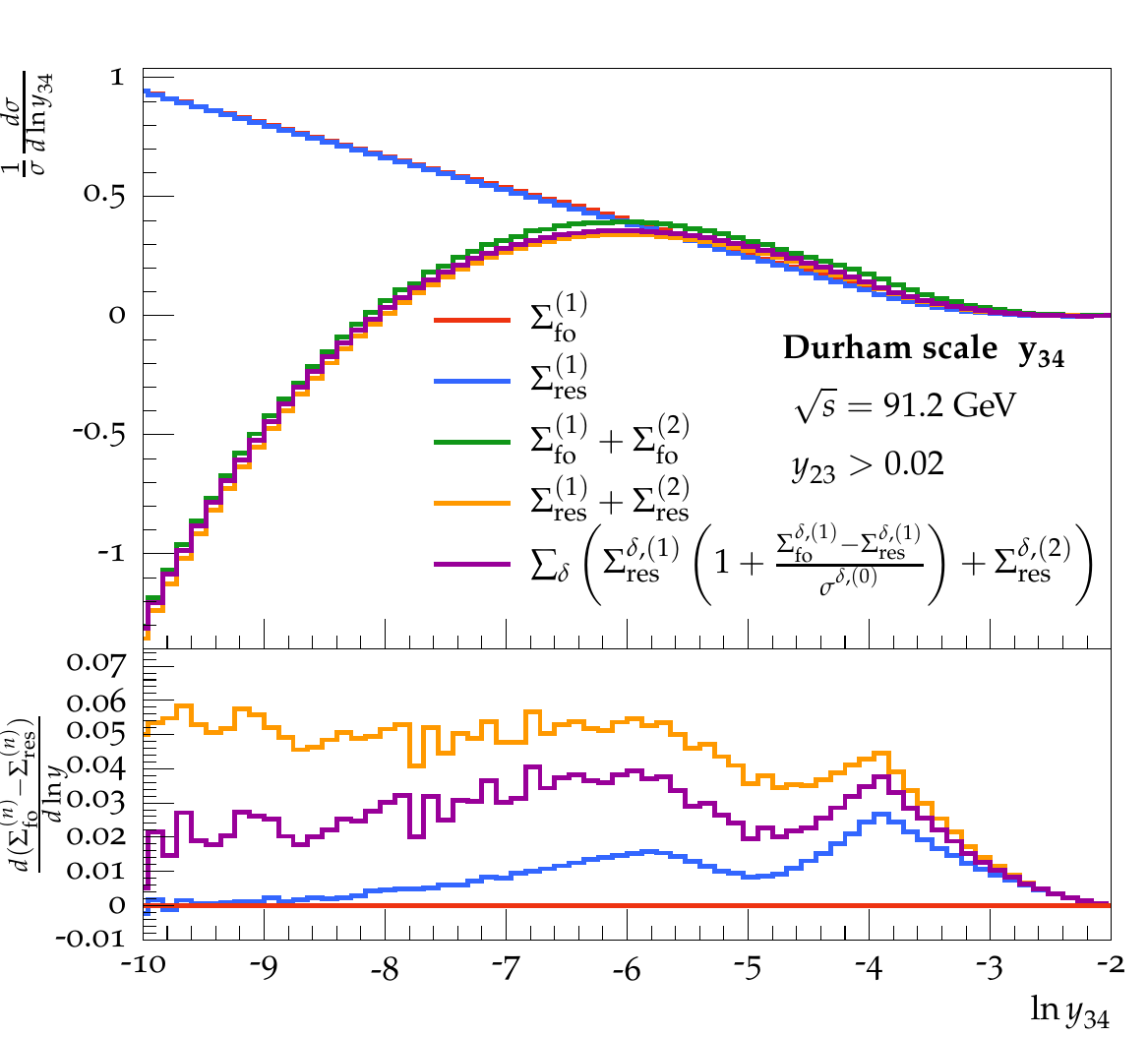}
  \includegraphics[width=.49\textwidth]{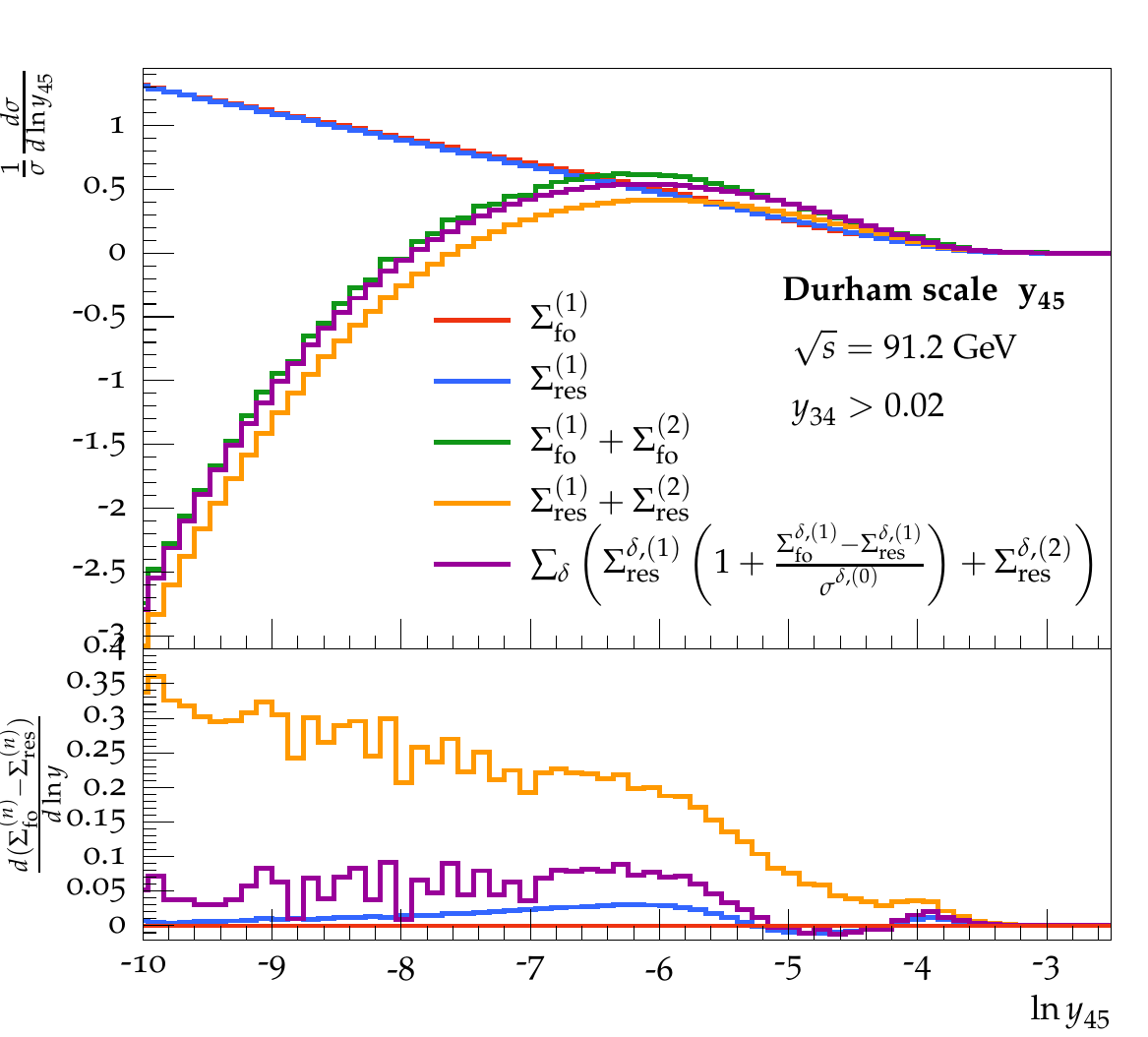}
  \includegraphics[width=.49\textwidth]{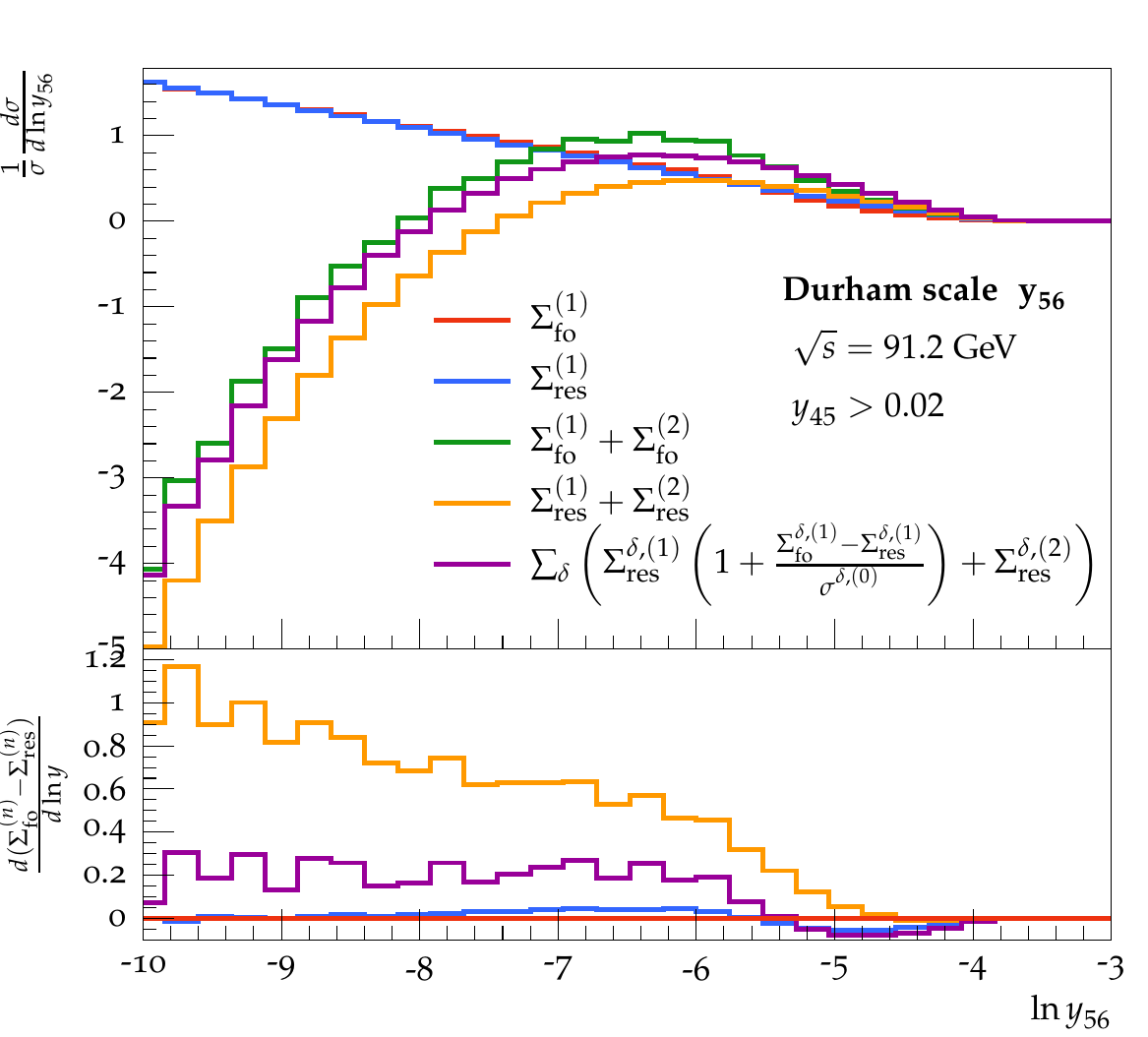}
  \caption{Fixed-order predictions for $y_{34}$,
    $y_{45}$, and $y_{56}$. We show the
    expansion of $\Sigma_\mathrm{res}$ to the relevant orders, and the expansion
    to second order including the expression approaching the $C_1$ coefficient
    in the soft limit. The lower panels show the total
    difference between fixed order and expansion.}\label{fig:fixed_order_results}
\end{figure}

In Fig.~\ref{fig:fixed_order_results}, we compare the fixed-order predictions
to the expansion obtained from the resummed results for the three observables
$y_{34}$, $y_{45}$, and $y_{56}$. We explicitly check their asymptotic agreement in
the limit $y_{n,n+1}\to 0$. As expected, the difference between fixed order and
the expansion of the resummed distribution approaches zero in the soft
limit. In the expansion $\Sigma^{(2)}$, there are missing terms of order
$\alphaS^{(n-2)}\alphaS^2L^2$ and $\alphaS^{(n-2)}\alphaS^2L$ that are present
in the fixed-order NLO calculation. Hence, we have a mismatch between fixed-order
calculation and expansion which, in the differential distribution, is growing
linearly in $L$, see the orange line in Fig.~\ref{fig:fixed_order_results}. After including the $\frac{\alphaS}{2\pi}C_1^\delta$
coefficients, which effectively happens in the matching, only missing terms of
order $\alphaS^{(n-2)}\alphaS^2L$  remain. These are subleading with
respect to the $\mathrm{NLL}^\prime$ resummation, although leading to a constant but
finite difference in the differential distributions between fixed order and
expansion. This is also demonstrated in Fig.~\ref{fig:fixed_order_results}, where in the
purple line we include the respective matching coefficient as it would effectively appear in
the multiplicative matching. As the NLO calcuation is computationally expensive,
we do not attempt to carry it out at sufficient precision to allow for an extraction
of the actual constant. Note that the matching schemes in
Sec.~\ref{sec:matching} are designed such that the matched distributions behave in a
physically meaningful way, i.e., are driven to zero by the resummed distribution,
cf.~Fig.~\ref{fig:resummed_results}, requiring a much less precise calculation in
the soft tail of the fixed-order correction.  
We compared our results presented in the previous section to the ones obtained
in the multiplicative matching scheme, but did not find significant differences,
except for the region $y_{n,n+1} > y_\mathrm{cut}$. There, however, scale uncertainties
also become very large, such that the difference between the two matching schemes
is always covered by the NLO scale variation in the LogR matching scheme. We thus
do not include an explicit uncertainty related to the matching scheme.  

\subsection{Impact of subleading colour contributions}\label{sec:impact_colour}

\begin{figure}[ht]
  \centering
  \includegraphics[width=.49\textwidth]{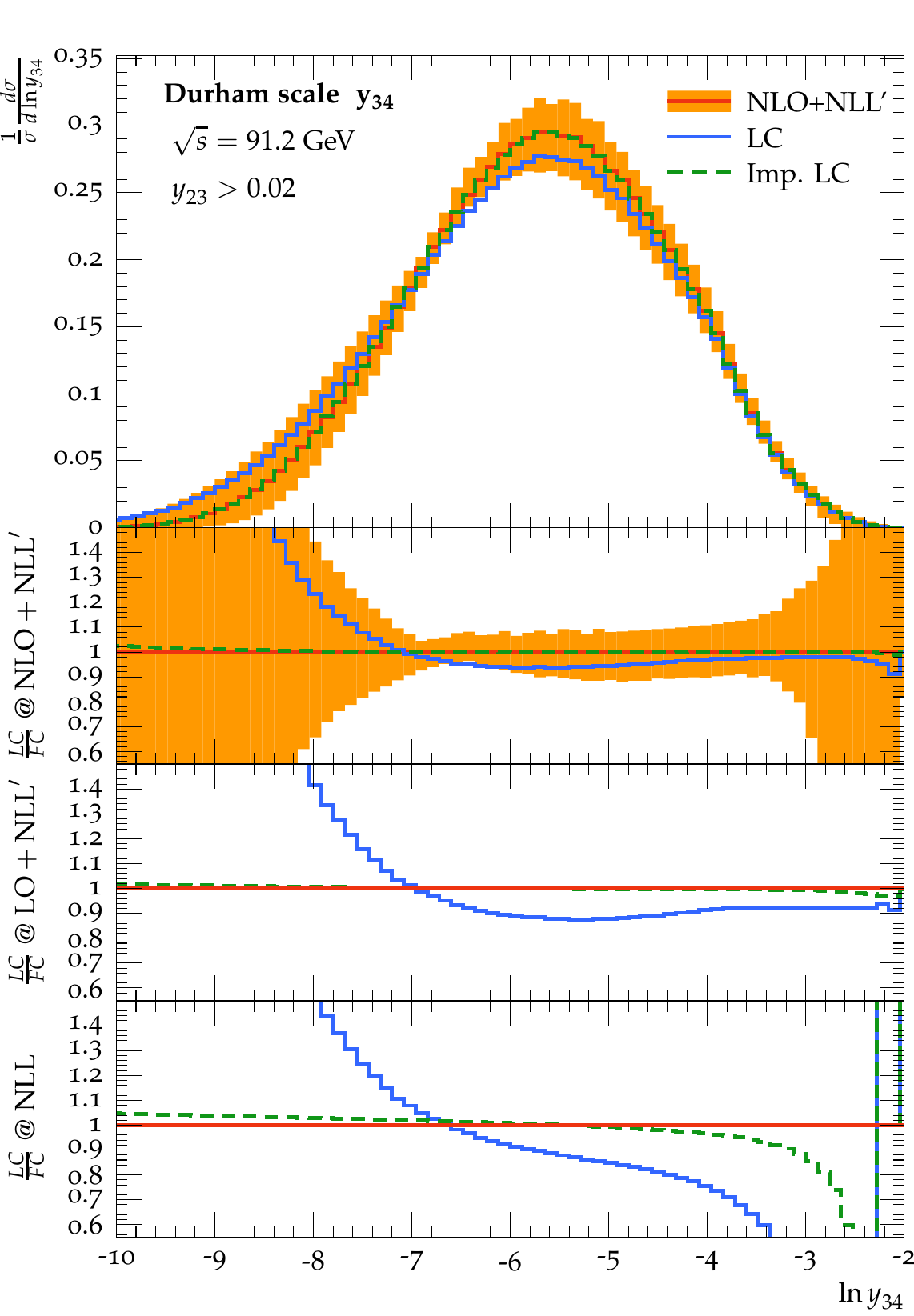}
  \includegraphics[width=.49\textwidth]{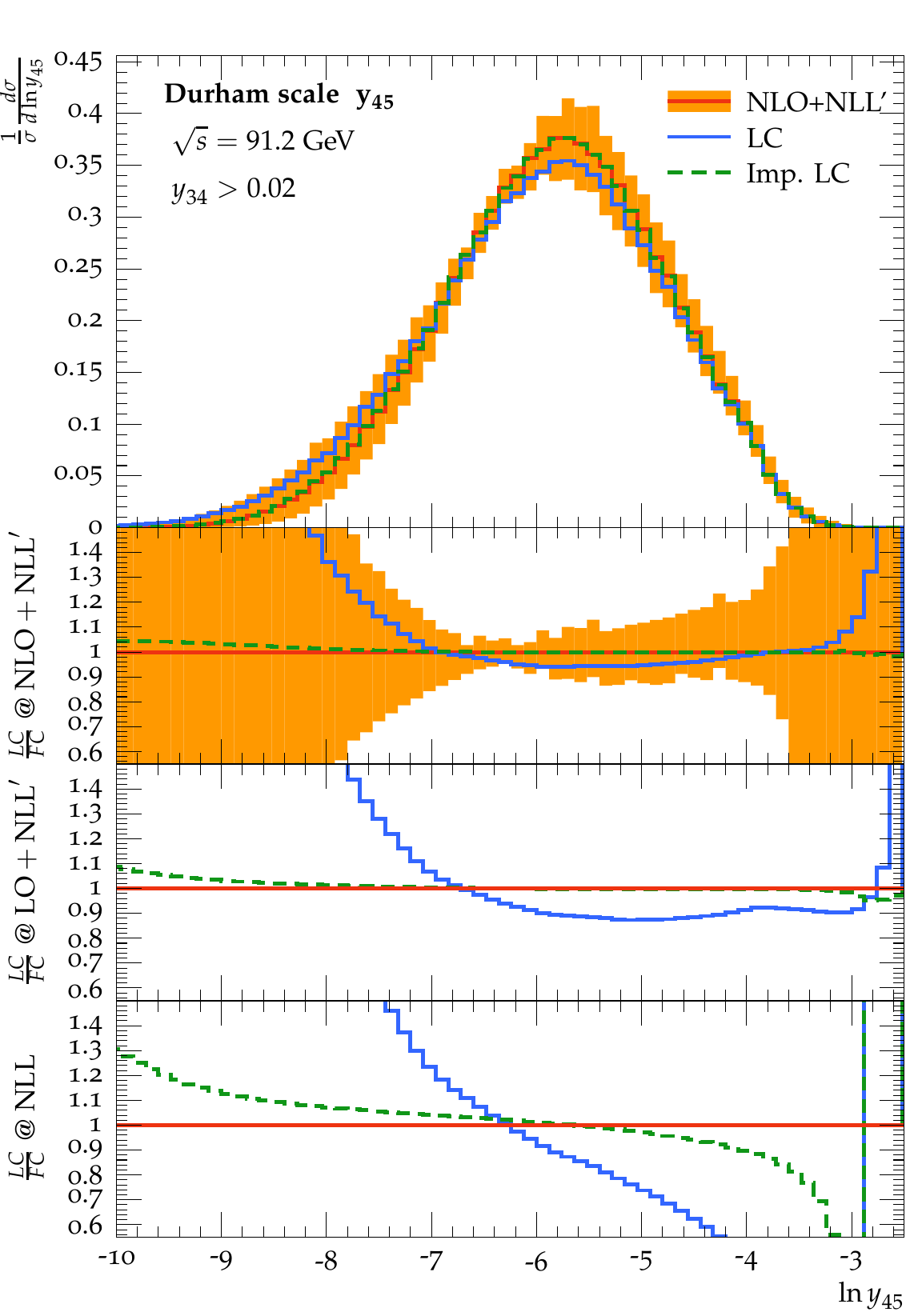}
  \includegraphics[width=.49\textwidth]{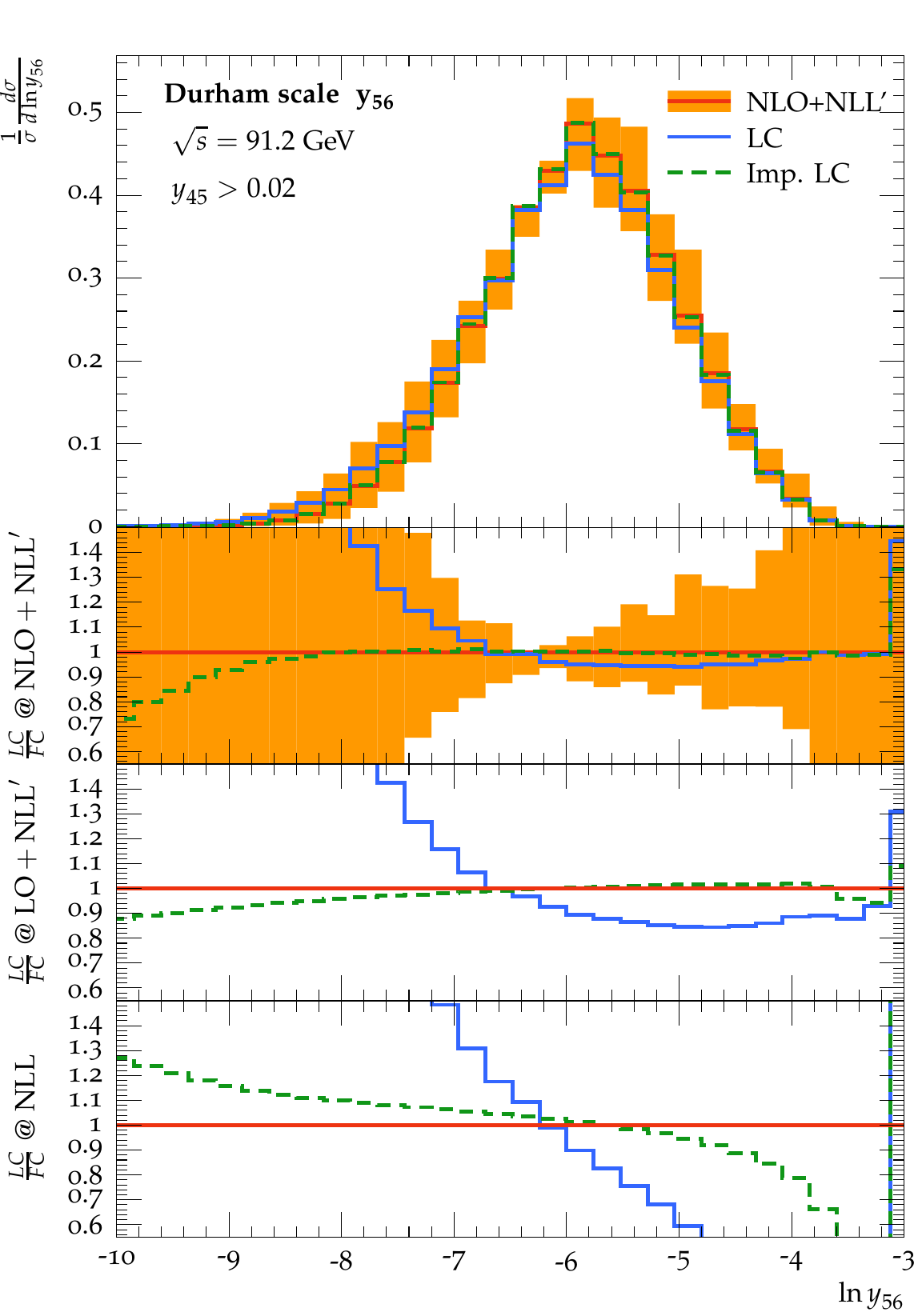}
  \caption{Durham splitting scales $y_{n,n+1}$ at $\mathrm{NLO+NLL}^\prime$ accuracy, in the
    $\Nc\to\infty$ limit, and in the improved LC scheme. The lower panels show
    the corresponding ratios additionally at $\mathrm{LO+NLL}^\prime$ accuracy
    and for the resummation without any matching.}%
  \label{fig:yij_LC}%
\end{figure}

Our results enable us to quantify the impact of subleading colour corrections on
multijet observables. This is of particular interest for example in the context
of recent and ongoing developments to systematically include such corrections
into parton-shower event generators, see
e.g.~\cite{Platzer:2012np,Platzer:2013fha,Platzer:2018pmd,Isaacson:2018zdi,Martinez:2018ffw,Forshaw:2019ver,Nagy:2019pjp,Nagy:2019rwb,Hoeche:2020nsx}. 
To assess the effect of subleading colour contributions, we redo our calculation
in the t'Hooft large-$\Nc$ limit, defined by taking $\Nc\to\infty$ while keeping $\alphaS \Nc$ fixed \cite{tHooft:1974pnl}. 
This approximation, to which we will refer to as leading colour (LC), has a significant
impact on the various contributions in Eq.~\eqref{eq:CAESAR}. 
Firstly, all Casimirs corresponding to different legs are simplified to $\Nc=2\CF=\CA$. 
Secondly, quark-loop contributions become negligible in both, the anomalous dimensions and the beta function. 
Finally, all non-planar diagrams vanish, leading to a simplification of colour-insertion operators. 
As the latter is only relevant to the colour-correlation contribution
$\mathcal{S}$, we define an "improved large-$\Nc$" approximation (imp. LC),
in which we only treat the colour correlators appearing in $\Gamma$, and hence
$\mathcal{S}$, in the strict $\Nc\to\infty$ limit, while still including the correct
subleading contributions in $R_l$ and $\mathcal{F}$. 
Those have a clear interpretation as sums over contributions from individual
legs, such that the proper leg-specific $\mathrm{SU(3)}$ Casimir operators can be assigned.
As in our main calculation, we match both resulting resummed distributions to
the full-colour NLO result in the LogR scheme.
The results of this study are presented in Fig.~\ref{fig:yij_LC} and we collect
details on the analytical treatment of subleading colour contributions in App.~\ref{app:formulas}. 

While there is a significant impact on the distributions when taking the t'Hooft
large-$\Nc$ limit, it is negligible for the improved large-$\Nc$ limit, although
in both cases, the results stay entirely within the uncertainty band of the
$\mathrm{NLO+NLL}^\prime$ prediction. Moreover, in the peak region there is
almost no difference between the full-colour and the improved large-$\Nc$
result; the only visible difference being in the soft-tail region, where,
however, the effect is below $5\%$ for $y_{34}$ and $y_{45}$ and only sizeable
in the ultra-soft region of $y_{56}$. 

To ease the interpretation of these results in the context of subleading colour corrections in
parton showers, which so far in the literature have mostly been discussed in the context of pure,
unmatched showers, we show the ratios between full, improved, and leading-colour approximation
at $\text{NLL}$, $\text{LO}+\text{NLL}^\prime$, and $\text{NLO}+\text{NLL}^\prime$ accuracy in
Fig.~\ref{fig:yij_LC}. It is evident that the effect of subleading colour contributions is largely captured by the
matching, which is expected as we include the full colour information in the fixed-order
calculations. Without any matching, we find the corrections to the improved leading-colour
scheme to be compatible with the naive expectation that, away from the kinematical extreme regions,
subleading colour contributions are suppressed by $1/\Nc^2 \approx 10\%$. Already LO matching
allows for a significant reduction of these effects. We remark, however, that a direct
interpretation of these observations in the context of full-colour parton showers is still not
straight forward, as the matching involves an intricate interleave of subleading colour and
kinematical corrections.  It is, however, reasonable to argue that subleading colour effects are
rather small and anyway overwhelmed by the fixed-order corrections. At least for the observables
considered here, there is virtually no difference between the calculations at full-colour and
in the improved large-$\Nc$ limit, as long as both are matched to exact, full-colour matrix elements.
  
\subsection{Comparison to parton-shower predictions}\label{sec:comparison_PS}

Our resummed $\mathrm{NLO+NLL}^\prime$ predictions for the Durham jet resolutions provide 
highly non-trivial benchmarks for corresponding predictions from QCD
Monte-Carlo event generators. They allow to gauge the results from
parton-shower simulations and the methods used to combine these with
exact higher-order QCD matrix elements and might guide the way to further
improving showering algorithms. On the other hand, Monte-Carlo simulations
provide means to include non-perturbative corrections from the parton-to-hadron
transition, that ultimately need to be taken into account for a realistic
comparison with experimental data. 

\begin{figure}%
  \centering
  \includegraphics[width=.49\textwidth]{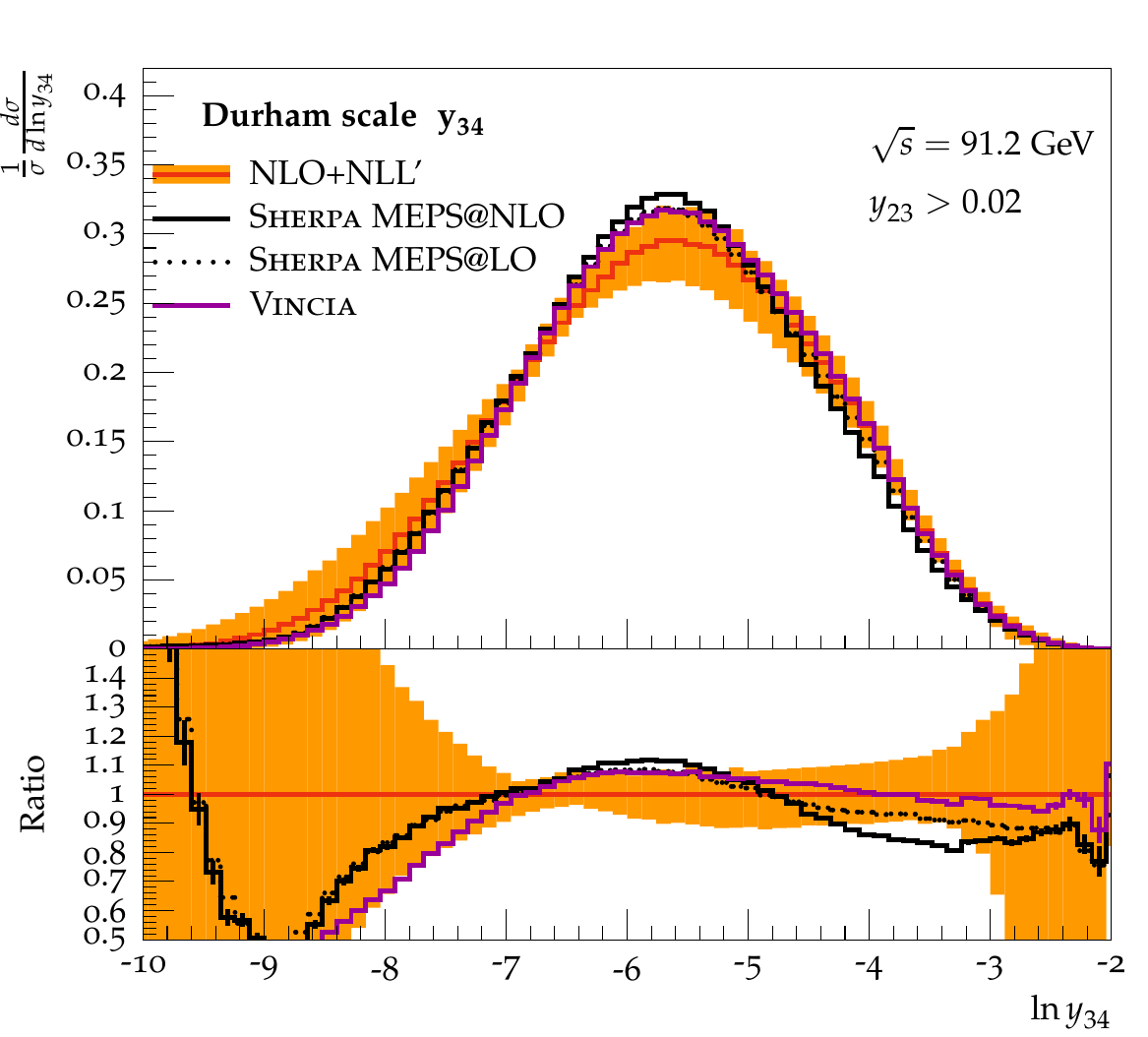}
  \includegraphics[width=.49\textwidth]{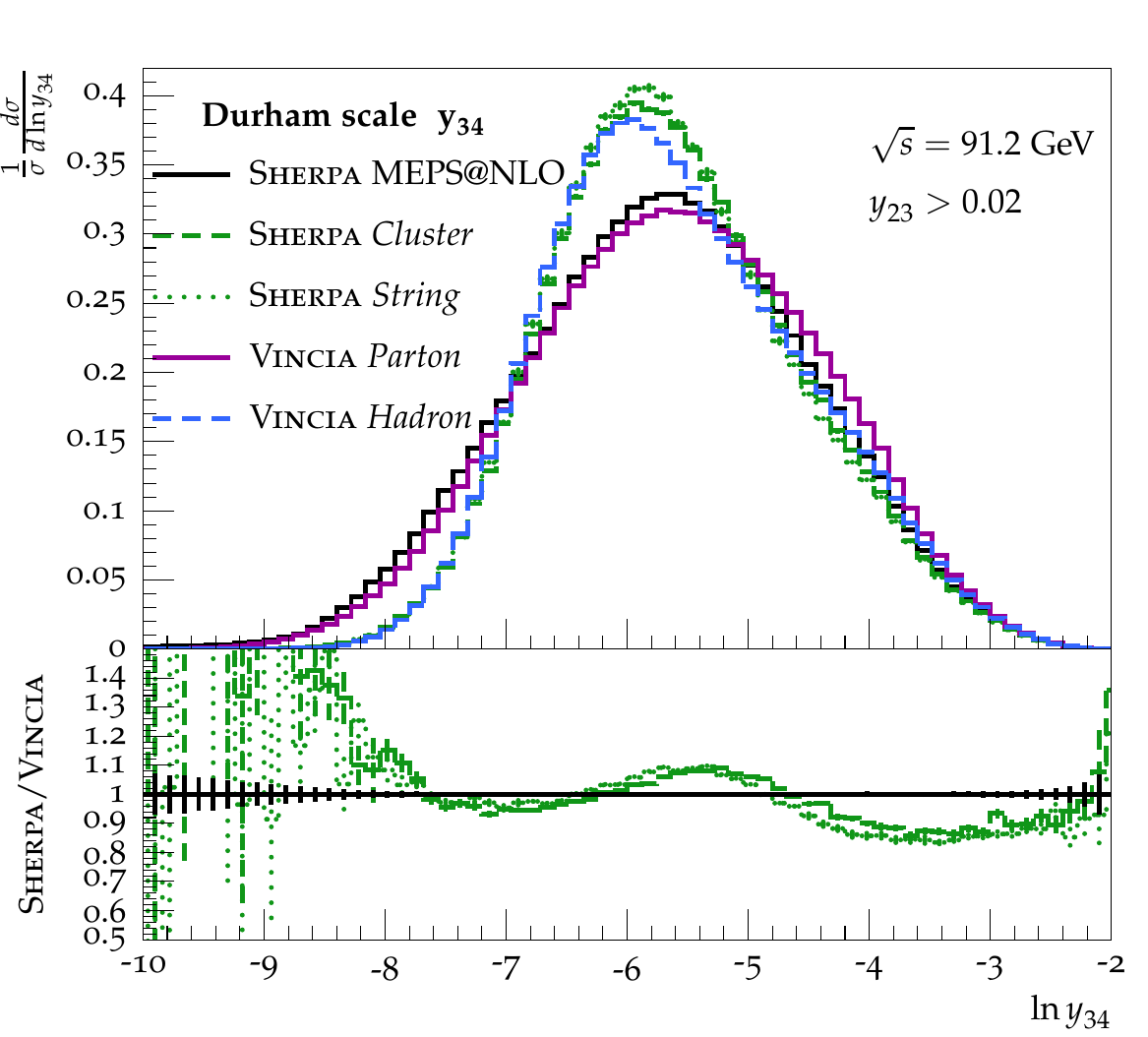}
  \includegraphics[width=.49\textwidth]{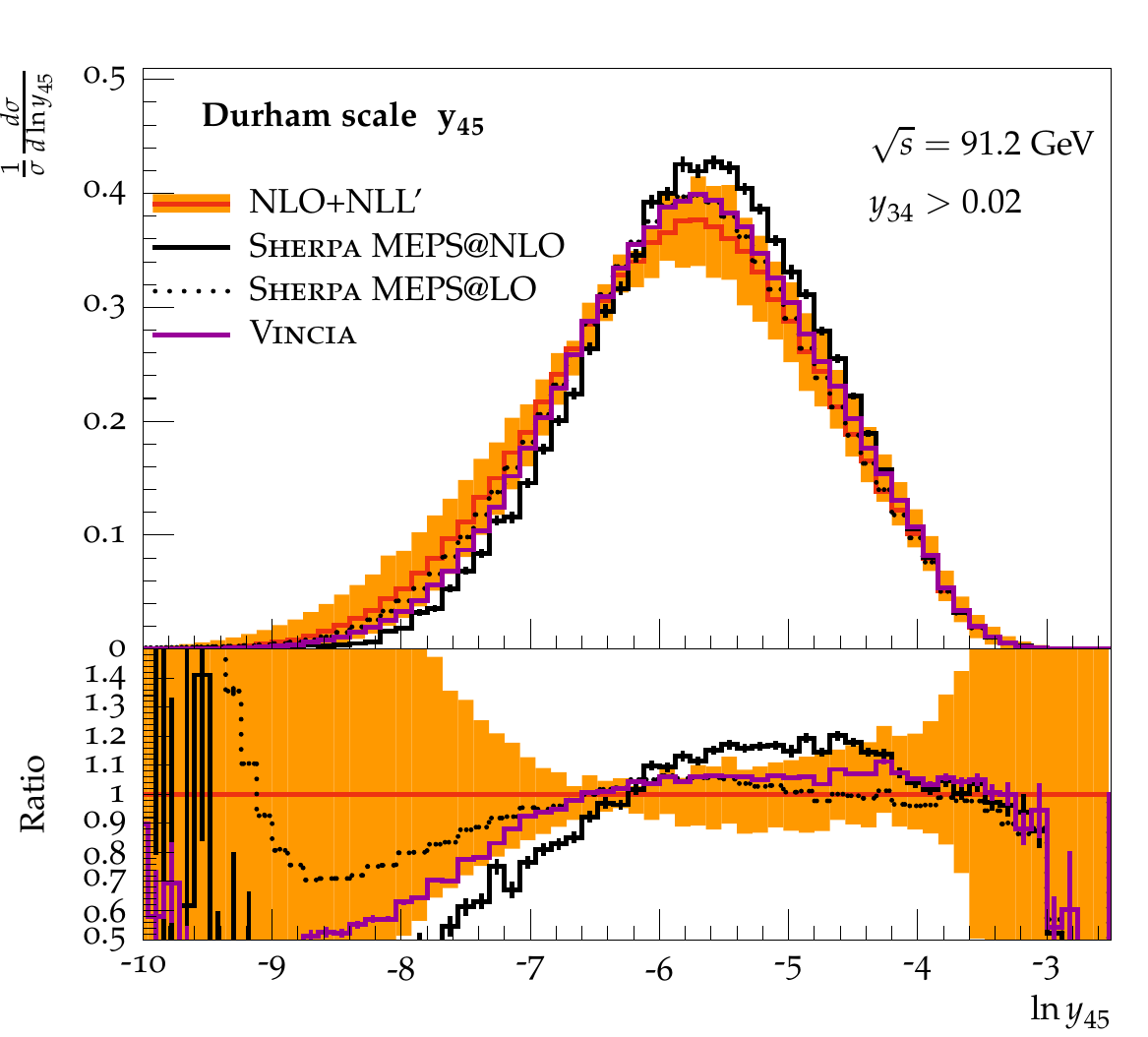}
  \includegraphics[width=.49\textwidth]{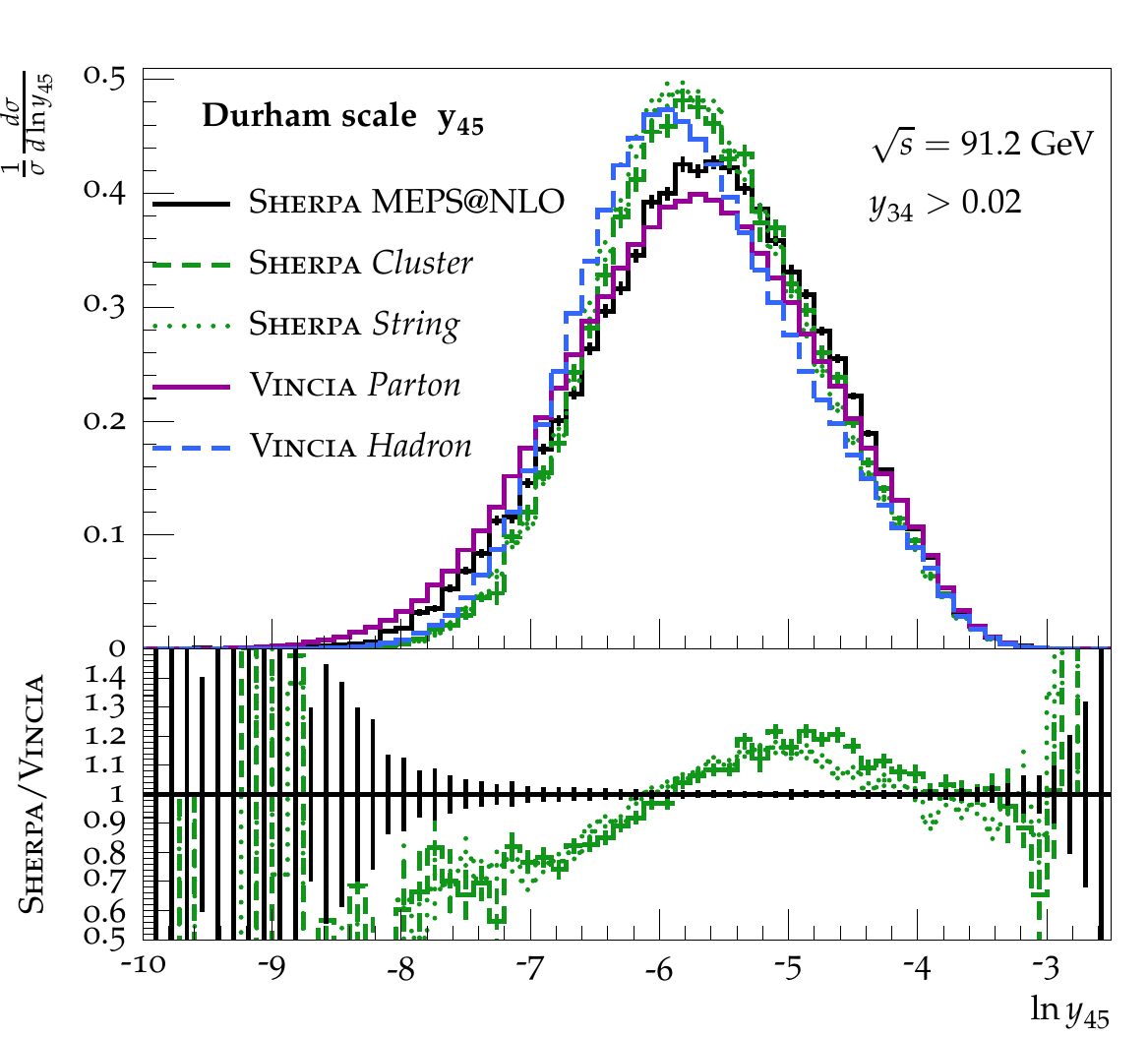}
  \includegraphics[width=.49\textwidth]{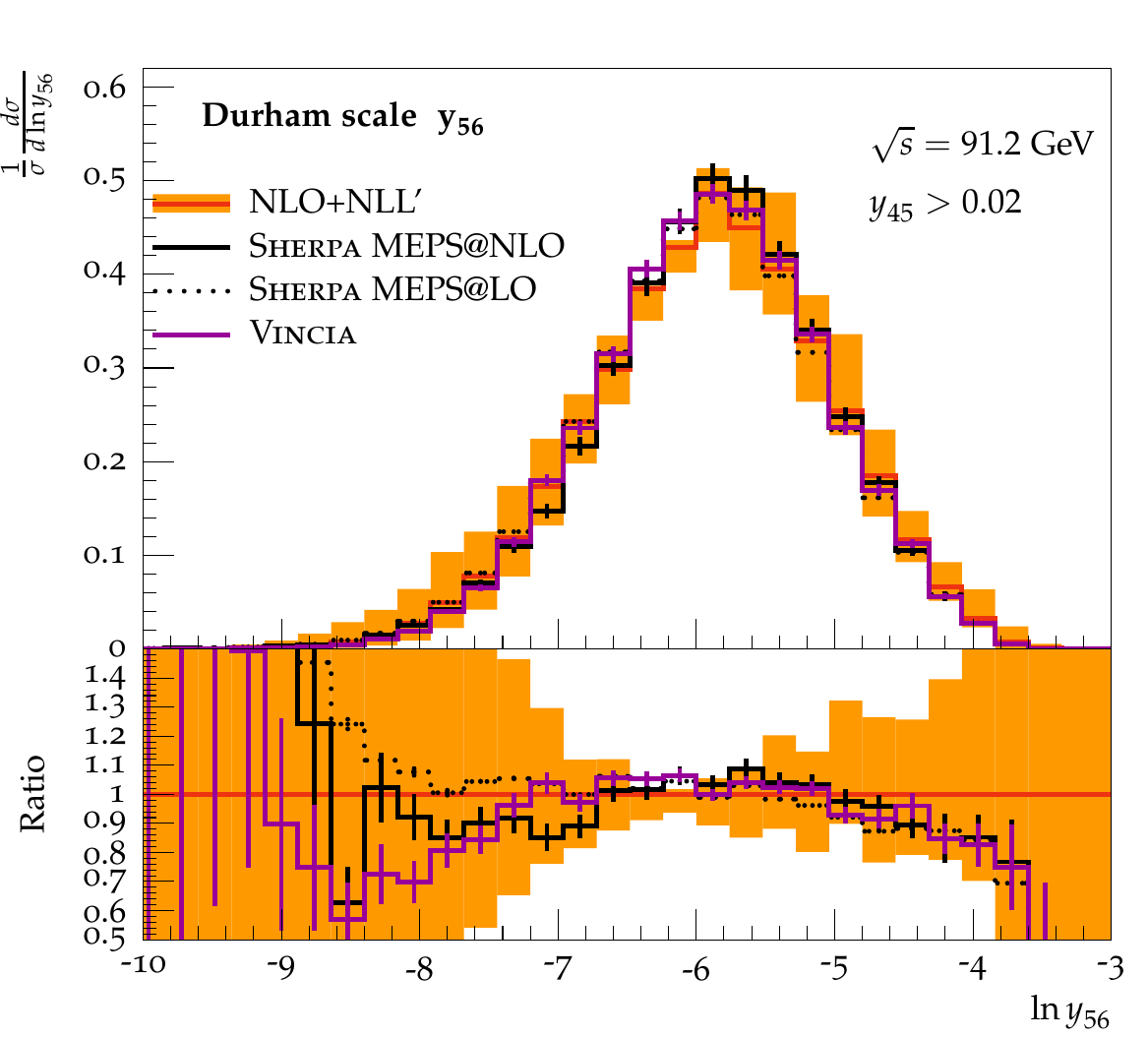}
  \includegraphics[width=.49\textwidth]{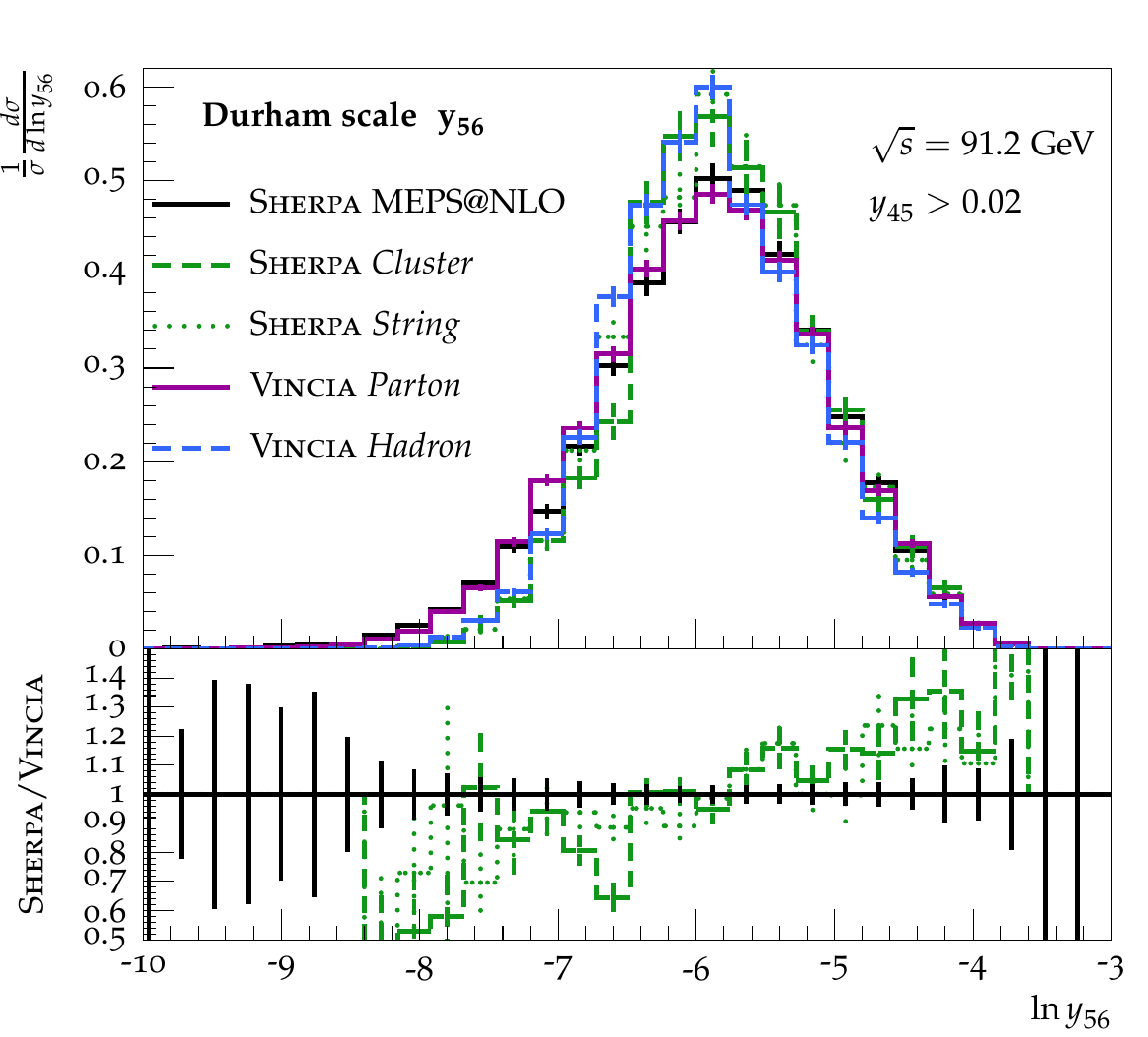}
  \caption{Comparison of resummed $\mathrm{NLO+NLL}^\prime$ predictions for $y_{34}$, $y_{45}$, and
    $y_{56}$ to parton-level parton-shower simulations from \Sherpa and \Vincia (left column).
    The right column compiles corresponding hadron-level results.
    The shower results from \Vincia are hadronised with the Lund string model
    in \Pythia 8. For the \CSS, the  \Sherpa implementation of the
    Cluster model is used (dashed line) as well as the Lund model from \Pythia 6
    (dotted line). The hadronisation corrections from \Sherpa are all based on the
    \MEPSatNLO\ sample.}%
  \label{fig:yij_PS}%
\end{figure}

As a first step in this direction we compare our resummed results against
predictions from two distinct parton-shower implementations: the dipole
parton shower \CSS~\cite{Schumann:2007mg} as implemented in the \Sherpa event
generator and the \Vincia\ 1 antenna-shower plug-in \cite{Giele:2007di}
to the \Pythia\ 8 event generator \cite{Sjostrand:2007gs}. In Fig.~\ref{fig:yij_PS}
we present our results obtained using matrix-element plus parton-shower
simulations at parton level for $y_{34}$, $y_{45}$, and $y_{56}$ in comparison
to the $\mathrm{NLO+NLL}^\prime$ predictions. Furthermore, in the right panels of
Fig.~\ref{fig:yij_PS} we present hadron-level predictions, compared to the
respective parton-level results.

To account for the kinematics and colour correlations of the respective Born
processes, both showers are corrected to exact LO and NLO calculations, using
two different strategies. One the one hand, in \Sherpa we consider an
\MEPSatLO~\cite{Hoeche:2009rj} calculation with exact tree-level matrix elements
for $e^+e^-\to 2,3,4,5$ final-state partons. We also compare to a calculation
using the \MEPSatNLO merging strategy~\cite{Hoeche:2012yf}, including one-loop
QCD corrections for $e^+e^-\to 2,3,4$ partons  via the \MCatNLO
method~\cite{Frixione:2002ik,Hoeche:2011fd} as implemented in \Sherpa and
$e^+e^- \to 5$ partons at tree level. We again make use of the matrix-element  generator
\Comix, and use \OpenLoops\ to compute the virtual corrections. The merging
parameter is set to 
$y_{\textrm{cut}}=\left(Q_{\textrm{cut}}/E_{\textrm{CMS}}\right)^2=10^{-2}$ for
the both the \MEPSatLO and the \MEPSatNLO calculation. 
The \Vincia antenna shower, on the other hand, is matched to $e^+e^- \to 2,3$
matrix elements at one-loop level as presented in \cite{Hartgring:2013jma} and
$e^+e^- \to 4,5,6$ tree-level matrix elements obtained from the \MadGraph\ 4
matrix-element generator \cite{Alwall:2007st} via matrix-element corrections in
the unitary GKS formalism \cite{Giele:2011cb}. Matrix-element corrections for
$5-$ and $6-$ parton processes are smoothly regularised at a matching scale of
$Q_\text{match}/E_\text{CMS} = 0.05$. For comparability, we restrict the phase space
to strongly ordered branchings only. In both showers, the strong coupling is evolved at two-loop order in the CMW
scheme, assuming an $\overline{\text{MS}}$ value of $\alphaS(m_Z^2)=0.118$. 

We observe that the \MEPSatLO and \Vincia samples agree well with each other and
are close to the analytic calculation in the peak region as well as in the hard region,
apart from the immediate neighbourhood of the endpoint. The \MEPSatNLO sample is
generally not improving the agreement between analytic and parton-shower
prediction. In fact for $y_{34}$ and $y_{45}$ it yields somewhat larger deviations.
However, we do not determine an explicit uncertainty estimate for the Monte Carlo
predictions here, but their size can be considered similar to those of the
analytic calculation. Accordingly, all the presented predictions are indeed in
very good agreement.  Comparing to our results in the previous section, the effects
of subleading colour contributions are qualitatively different and smaller than the
differences we observe between our resummed results and the parton-shower predictions.
Apparently, for the particular observable definition we consider here, missing subleading
colour contributions are not driving these differences but rather ambiguities related
to recoil schemes, phase-space constraints and the treatment of subleading
contributions in the running of $\alphaS$, see for example
Refs.~\cite{Hoeche:2017jsi,Dasgupta:2018nvj,Bewick:2019rbu}.

As our observable definition differs from the usual experimental definition of
jet-resolution scales by the hard cut we impose on the Born event, we need to
gauge the influence of the parton-to-hadron transition on this observable. To this
end we compare generator predictions on parton and hadron level. For \Sherpa, we invoke
\Sherpa's default hadronisation model, based on cluster
fragmentation~\cite{Winter:2003tt} and furthermore, hadronise \Sherpa's
parton-level events with the Lund string fragmentation~\cite{Sjostrand:1982fn}
as implemented in \Pythia\ 6.4~\cite{Sjostrand:2006za}. Parton-level predictions
of the \Vincia antenna shower are hadronised using the Lund string model in
\Pythia~8.2.

We obtain sizeable hadronisation corrections in all multiplicities. Their impact is
very similar for the cluster- and string-fragmentation model applied to \Sherpa's
parton-level results. A qualitatively similar effect can be seen in the string
hadronisation of \Vincia's parton-level results. In all cases, the soft tail of the
distributions is significantly suppressed, leading to a narrowing of the distribution
and hence a more pronounced peak. Reassuringly, the hard tail is only mildly
affected by either of the fragmentation models. Quantitatively, however,
there are differences of up to $10-20\%$ remaining between the hadron-level
predictions from \Sherpa and \Vincia in the central peak region. We note that this
is largely compatible with the deviations seen in the comparison at parton level.
As the two predictions from \Sherpa with different hadronisation models agree 
relatively well, it can be expected that generic non-perturbative uncertainties are
rather moderate. Moreover, taking into account that we do not include uncertainties
regarding the variation of hadronisation parameters, all hadron-level predictions
agree reasonably well with each other. 

\section{Conclusions}\label{sec:conclusions}

For the first time, we have here obtained resummed predictions at $\mathrm{NLO+NLL}^\prime$ accuracy
for multijet resolution scales in electron-positron annihilation. We
employ the \Caesar\ formalism in a largely automated manner
within the \Sherpa event-generator framework. All relevant colour spaces were decomposed over the trace basis to obtain
hard-scattering matrix elements and colour correlators that account for the insertion of soft-gluon
radiation. Both, the construction of the basis and the calculation of colour insertions, have been performed in an automated way.
Multijet matrix elements are obtained from the \Comix matrix-element
generator, that is part of \Sherpa. For NLO QCD predictions we obtain virtual
corrections for $e^+e^-\to 3,4,5$ partons from \OpenLoops, while using \Recola\ for virtual corrections to $e^+e^-\to 6$ partons. For the evaluation of the multiple-emission contribution, represented by the $\mathcal{F}$-function in the \Caesar formalism, we resort to a numerical
evaluation using multiple-precision arithmetics.

We have derived predictions for the Durham jet-resolution scales $y_{34}$, $y_{45}$, and
$y_{56}$ at $\mathrm{NLO+NLL}^\prime$ accuracy, using a Durham resolution $y_{\mathrm{cut}} = 0.02$ to restrict
the respective Born configurations to sufficiently hard kinematics. The inclusion
of NLO QCD corrections significantly reduces the theoretical uncertainties, estimated
via scale variations. 

We studied the impact of subleading colour contributions to our predictions, by repeating
our calculations in the LC as well as the improved LC scheme, which we regard as being similar to the colour
treatment in parton showers. We observe significant differences to the full-colour prediction only in
the (strict) LC limit, while already at NLL the improved LC scheme well approximates the
full-colour prediction. At $\mathrm{NLO+NLL}^\prime$ accuracy, we observe virtually no difference between
the improved LC and the full-colour result.

As a benchmark for parton showers and an estimate of non-perturbative effects on the observable, we compared
our resummed predictions against two distinctly different parton-shower algorithms, the \Vincia\ antenna shower
plug-in to \Pythia\ and the Catani--Seymour dipole shower in \Sherpa, matched to LO and NLO. We observe good
agreement between the \Vincia\ and \MEPSatLO\ as well as the \MEPSatNLO\ results. The found effects of subleading colour contributions are qualitatively different and smaller than the
differences observed when comparing resummed results and parton-shower predictions. Apparently these
will rather originate from ambiguities related to recoil schemes, phase-space constraints and the
treatment of subleading contributions in the running of $\alphaS$ in the parton-shower implementations.
The effect of non-perturbative corrections was studied by including hadronisation effects for
the \Vincia\ and \Sherpa\ parton showers, employing both, cluster and string fragmentation for the
latter. We found good agreement between all hadron-level results. The observed deviations between
\Sherpa and \Vincia can already be found at the parton level. This confirms the suitability of
jet-resolution scales for studies of perturbative QCD dynamics.

A direct comparison of our predictions to existing LEP measurements of jet resolutions is not straight-forward, due to the required regularisation of the multijet Born processes. A corresponding reanalysis of LEP data would be desirable. To be able to compare to existing $e^+e^-$ data, more general hierarchies of the multi-scale problem need to be addressed. By the generality of the approach presented here, our study may be extended to $k_T$ jet-resolution scales in hadronic collisions, motivating dedicated measurements at the LHC.

\section*{Acknowledgements}
We would like to thank Stefan H\"oche and Vincent Theeuwes for useful discussions. CTP
thanks Peter Skands for support. Our work has received funding from the European Union’s
Horizon 2020 research and innovation programme as part of the Marie Sk\l{}odowska-Curie
Innovative Training Network MCnetITN3 (grant agreement no. 722104). SS acknowledges support
through the Fulbright-Cottrell Award and from BMBF (contract 05H18MGCA1).
DR acknowledges support from the German-American
Fulbright Commission allowing him to stay at Fermi National Accelerator Laboratory (Fermilab),
a U.S. Department of Energy, Office of Science, HEP User Facility. Fermilab is
managed by Fermi Research Alliance, LLC (FRA), acting under Contract No.\
DE--AC02--07CH11359. 
CTP is supported by the Monash Graduate Scholarship, the Monash International Postgraduate Research Scholarship, and the J. L. William Scholarship.
CTP is also grateful to
Fermilab for hospitality and travel support.
This work was
further partly funded by the Australian Research Council via Discovery Project
DP170100708 – “Emergent Phenomena in Quantum Chromodynamics”.

\appendix
\section{Analytic expressions}\label{app:formulas}
For completeness, we here collect all analytic expressions of the \Caesar
formalism that did 
not appear in the main text, restricted to the jet-resolution scales
studied here and our conventions. In particular, we choose to rescale the
arguments of all logs with $x_v = d(\mu_Q)$, so that no explicit dependence on
$d$ appears. We use the short-hand $\lambda = \alphaS(\mu_R^2)\beta_0 L$, and fix
the normalisation of colour operators to $\TR=1/2$, resulting in $C_l = \CA=\Nc$
for gluons and $C_l =\CF=(\Nc-1/\Nc)/2$ for quarks. The radiator of leg $l$ is
given by 
\begin{equation}
  \begin{split}
  R_l(L) &= -\frac{C_l}{2\pi \beta_0}\left[L\left(1 + \frac{\ln(1-\lambda)}{\alphaS
      \beta_0}\right)\right. -  \left(2 B_l+\ln\frac{Q^2}{\mu_Q^2}\right) \ln(1-\lambda) \\
 &\left.+
 \left(\frac{K}{2\pi\beta_0}+\ln\frac{\mu_R^2}{\mu_Q^2}\right)\left(\ln(1-\lambda)+\frac{\lambda}{1-\lambda}\right)+\frac{\beta_1}{\beta_0^2}\left(\frac{1}{2}\ln^2(1-\lambda)+\frac{\ln(1-\lambda)+\lambda}{1-\lambda}\right)\right]\,.
 \end{split}
\end{equation}
Its derivative with respect to $L$ is single logarithmic,
\begin{equation}\label{eq:Rprime}
  R^\prime_l = \frac{C_l}{2\pi\beta_0} \frac{\lambda}{1-\lambda}.
\end{equation}
We also use the shorthands $R = \sum_l R_l$, $R^\prime = \sum_l
R^\prime_l$ where this eases the notation. The usual constants in the
$\overline{\text{MS}}$ scheme are
\begin{equation}
  \begin{split}
    \beta_0 &= \frac{11\CA-2\nf}{12 \pi}\,,~& 
    \beta_1 &= \frac{17\CA^2-5\CA\nf-3\CF\nf}{24\pi^2}\,,\\ 
    K &= \CA \left(\frac{67}{18}-\frac{\pi^2}{6}\right) - \frac{5}{9}\nf\,,~&B_q &= -\frac{3}{4},~&B_g &= -\frac{\pi\beta_0}{\CA}\,,
  \end{split}
\end{equation}
where we have already employed our normalisation convention. In the t'Hooft limit,
$\Nc\to\infty$ with $\alphaS \Nc = \alphaSzero = \mathrm{const.}$, all
expressions depend only on the finite quantities 
\begin{equation}
  \begin{split}
    \frac{\CF}{2\pi\beta_0}&\to \frac{3}{11}\,,~&\frac{\CA}{2\pi\beta_0} &\to
    \frac{6}{11}\,,~&\alphaS\beta_0 &\to \frac{11}{12\pi}\alphaSzero\,,\\
    \frac{K}{2\pi\beta_0} &\to \frac{67/3-\pi^2}{11}\,,~&  \frac{\beta_1}{\beta_0^2} &\to \frac{102}{121}\,.
  \end{split}
\end{equation}
When working in the large-$\Nc$ limit, the $\mathcal{F}$ function needs to be re-evaluated in principle.
However, it only depends on the ratios of sums of Casimirs for the different legs and on
$R^\prime$. Thus only the configurations with mixed quark and gluon content need to be re-computed
with $\CA/\CF=2$.

When evaluating colour correlators in the large-$\Nc$ limit, we are interested in
\begin{equation}
  \alphaS \frac{\bra{b_\alpha}\boldsymbol T_i \boldsymbol
    T_j\ket{b_\beta}}{\vert b_\alpha \vert \vert b_\beta \vert} \to \alphaSzero
  \bra{b_\alpha}\boldsymbol T_i \boldsymbol T_j\ket{b_\beta}_{\mathrm{LNC}}\,,
  \hspace{1cm} \bra{b_\alpha}\boldsymbol T_i \boldsymbol T_j\ket{b_\beta}_{\mathrm{LNC}} = \lim_{\Nc \to \infty} \frac{\bra{b_\alpha}\boldsymbol T_i \boldsymbol T_j\ket{b_\beta}}{\Nc \vert b_\alpha \vert \vert b_\beta \vert}\,. \label{eq:large_NC_correlator}
\end{equation}
Note that we need to normalise the basis vectors in order to obtain a finite result. As
expected, the $\bra{b_\alpha}\boldsymbol T_i \boldsymbol T_j\ket{b_\beta}_{\mathrm{LNC}}$ correlators
vanish for all non-planar diagrams and give finite contributions
otherwise. Practically, we calculate the large-$\Nc$ colour correlators by comparing the powers of $\Nc$ in the
numerator and denominator of Eq.~\eqref{eq:large_NC_correlator} and keeping only
those with the same highest power. This modification of the colour correlators
is the only one that is still present in the improved LC scheme.

\bibliographystyle{amsunsrt_modp}
\bibliography{rjr}

\end{document}

%% file: preample.tex
\usepackage{authblk}

\usepackage{booktabs}
\usepackage{todonotes}
\presetkeys{todonotes}{inline}{}
\usepackage{amsmath}
\usepackage{amssymb}
\usepackage{commath}
\usepackage{array}
\usepackage{calc}
\usepackage{longtable}
\usepackage{multirow}
\usepackage{physics}
\usepackage{tensor}
\usepackage{upgreek}
\usepackage{pstricks}
\usepackage{graphicx}
\graphicspath{{figures/}}
\usepackage{xspace}
\usepackage{listings}
\usepackage[section]{placeins}
\usepackage{
  pgf,
  tikz}
\usetikzlibrary{
  patterns,
  shapes.multipart,
  arrows,
  trees,
  scopes,
  decorations.pathreplacing,
  decorations.pathmorphing,
  decorations.markings,
  decorations.text,
  positioning,
  calc
}

\usepackage[utf8]{inputenc}
\usepackage{mciteplus}
\usepackage[a4paper,pdfborder={0 0 0}]{hyperref}
\usepackage[format=hang,labelfont=bf,hypcap=true]{caption}
\usepackage{subfig}
\usepackage{sectsty}
\usepackage{relsize}

\allsectionsfont{\sffamily}
\subsubsectionfont{\mdseries\itshape\large}

\let\spreprint\empty
\newcommand{\preprint}[1]{\def\spreprint{\protect#1}}
\let\sinstitute\empty

\makeatletter
\renewcommand{\maketitle}{\begingroup
  \null\thispagestyle{empty}%
    \ifx\spreprint\empty
      \vskip 5ex
    \else
      \flushright\large\spreprint\vskip 2ex
    \fi
    \vskip 5ex
    \flushleft
      {\sffamily\bfseries\huge\@title}\vskip 2ex
      \@author\vskip 2ex
      \ifx\sinstitute\empty
      \else
        {\small\sinstitute}
      \fi
    \vskip 5ex
  \endgroup
}
\makeatother
\renewenvironment{abstract}{\begin{center}
  {\large\sffamily\bfseries Abstract: }
  \begin{minipage}[t]{0.75\textwidth}
}{\end{minipage}\end{center}\vskip 10ex}

\newcommand{\Sherpa}{S\protect\scalebox{0.8}{HERPA}\xspace}
\newcommand{\Pythia}{P\protect\scalebox{0.8}{YTHIA}\xspace}

\newcommand{\Comix}{C\protect\scalebox{0.8}{OMIX}\xspace}
\newcommand{\CSS}{CSS\protect\scalebox{0.8}{HOWER}\xspace}
\newcommand{\Caesar}{C\protect\scalebox{0.8}{AESAR}\xspace}
\newcommand{\Vincia}{V\protect\scalebox{0.8}{INCIA}\xspace}
\newcommand{\MCatNLO}{MC\protect\scalebox{0.8}{@}NLO\xspace}
\newcommand{\MEPSatLO}{MEPS\protect\scalebox{0.8}{@}LO\xspace}
\newcommand{\MEPSatNLO}{MEPS\protect\scalebox{0.8}{@}NLO\xspace}
\newcommand{\ColorFull}{C\protect\scalebox{0.8}{OLOR}F\protect\scalebox{0.8}{ULL}\xspace}
\newcommand{\MadGraph}{M\protect\scalebox{0.8}{AD}G\protect\scalebox{0.8}{RAPH}\xspace}
\newcommand{\OpenLoops}{\textsc{OpenLoops}}
\newcommand{\Recola}{\textsc{Recola}}
